\documentclass[twocolumn,letterpaper]{IEEEAerospaceCLS}  

\usepackage[]{graphicx}    

\newcommand{\ignore}[1]{}  

\usepackage{xurl} 
\usepackage{amsmath,amsfonts,bm}
\DeclareMathOperator*{\argmin}{argmin} 

\usepackage{cite}

\usepackage{subfigure} 
\usepackage{cleveref}

\usepackage{xcolor}
\newcommand{\todo}[1]{\textcolor{red}{\textbf{\underline{TODO:}} #1}}
\renewcommand{\todo}[1]{}  
\usepackage{float}

\begin{document}
\title{Run Time Assurance for Simultaneous Constraint Satisfaction During Spacecraft Attitude Maneuvering\thanks{Approved for public release; distribution is unlimited. Case Number: AFRL-2023-4300.}}

\author{
Cassie-Kay McQuinn\\ 
Department of Aerospace Engineering\\
Texas A\&M University\\
College Station, TX, 77843\\
ckmcquinn@tamu.edu
\and 
Kyle Dunlap\\
Parallax Advanced Research\\
4035 Colonel Glenn Hwy\\
Beavercreek, OH, 45431 \\
kyle.dunlap@parallaxresearch.org
\and 
Nathaniel Hamilton\\
Parallax Advanced Research\\
4035 Colonel Glenn Hwy\\
Beavercreek, OH, 45431 \\
nathaniel.hamilton@parallaxresearch.org
\and
Jabari Wilson\\
Department of Mechanical Engineering\\
University of Florida\\
Gainesville, Florida, 32611\\
jabari.wilson@outlook.com
\and
Kerianne L. Hobbs\\ 
Air Force Research Laboratory\\
2241 Avionics Circle\\
Wright-Patterson AFB, OH, 45433\\
kerianne.hobbs@us.af.mil
\thanks{\footnotesize 979-8-3503-0462-6/24/$\$31.00$ \copyright2024 IEEE}              
}

\maketitle

\thispagestyle{plain}
\pagestyle{plain}

\maketitle

\thispagestyle{plain}
\pagestyle{plain}

\begin{abstract}

 A fundamental capability for On-orbit Servicing, Assembly, and Manufacturing (OSAM) is inspection of the vehicle to be serviced, or the structure being assembled. This research assumes autonomous slewing to maintain situational awareness of multiple vehicles operating in close proximity where several safety constraints must be satisfied. A variety of techniques may be used as the primary controller. 
 The focus of this research is developing Run Time Assurance (RTA) filters that monitor system behavior and the output of the primary controller 
 to enforce safety constraint satisfaction. 
Specifically, this research explores combining a subset of the constraints into an Active Set Invariance Filter (ASIF) RTA defined using control barrier functions. This method is minimally invasive to the primary control by minimizing deviation from the desired control output of the primary controller, 
while simultaneously enforcing all safety constraints. The RTA is designed to ensure the spacecraft maintains attitude requirements for communication and data transfer with a ground station during scheduled communication windows, adheres to conical attitude keep out zones, limits thermally unfavorable attitude duration, maintains attitude requirements for sufficient power generation, ensures maneuvers are below threshold to cause structural damage, ensures maximum angular velocity is below limits to maintain ability to respond quickly to new slewing commands, and conserves actuator use to prevent wear when possible. Slack variables are introduced into the ASIF controller to prioritize safety constraints when a solution to all safety constraints is infeasible. Monte Carlo simulation results as well as plots of example cases are shown and evaluated for a three degree of freedom spacecraft with reaction wheel attitude control. 

\end{abstract}

\tableofcontents

\section{Introduction}


According to the United Nation's Register of Objects Launched into Outer Space \cite{OSOidx}, 2474 registered objects were launched into outer space in 2022. Compared to the 222 registered objects that were launched in 2015, the space industry is growing at a rapid pace. With this growth comes a change in the manner with which missions are conducted, introducing a paradigm shift with the emerging technology of On-orbit Servicing, Assembly, and Manufacturing (OSAM) \cite{arney2021orbit}. OSAM will facilitate extended space missions, orbital manipulation of existing objects, such as the removal of space debris, and repair of existing satellite systems. A foundational capability of OSAM missions is the inspection task. Inspection is the observation of systems to analyze features of interest, including damaged assets or assembly of new components. Inspection vehicles must maintain situational awareness of multiple vehicles operating in close proximity. Additionally, safety and mission constraints on the inspection vehicle must be satisfied for the vehicle to successfully complete the inspection task. 

Introducing autonomy into spacecraft operations has the opportunity to minimize errors, human performance limitations, and operational costs. However, in order to utilize autonomous systems, safety of the system must be guaranteed. Specifically for the inspection task, unsafe controls could result in collision or loss of control causing damage to the satellite or system under inspection. This could cause loss of multi-million dollar systems as well as create space debris which can impact other spacecraft operations.


Previous research has used hazard analysis methods to elicit safety requirements for autonomous spacecraft maneuvering \cite{hobbs2021risk}. These specifications are formalized mathematically in this manuscript to define a safe set of states that adhere to all the attitude safety constraints. However, defining a single controller that can adhere to multiple simultaneous safety constraints while balancing performance is challenging. This complex controller design and verification challenge has led to the development of an online safety verification technique called \textit{run time assurance} (RTA), that creates a filter designed specifically to monitor the system and primary controller to ensure adherence to safety constraints \cite{hobbs2023runtime}. Even with this separation of performance and safety objectives within a control system, deriving a controller that keeps the spacecraft within a safe set of multiple safety constraints for all time is challenging using traditional control techniques. One approach to RTA for multiple safety constraints is to design monitors and backup controllers for each individual safety constraint as proposed in the Expandable Variable Autonomy Architecture (EVAA) RTA architecture \cite{hook2018initial}. However, these multi-monitor approaches prioritize a single safety constraint over others when multiple safety constraints may be violated. Instead, building off of the work by Dunlap et al. \cite{dunlap2023run,dunlap2023run2}, this research utilizes an RTA approach based on Active Set Invariance Filtering (ASIF) that simultaneously satisfies multiple safety constraints, defined as control barrier functions (CBFs). 

CBFs have shown to be an effective method for safety assurance of autonomous systems for automotive systems \cite{ames2016control}, robotic hand control in the presence of mode uncertainty \cite{cortez2019control}, bipedal robots \cite{hsu2015control}, spacecraft docking \cite{dunlap2021comparing}, and spacecraft translational maneuvering \cite{dunlap2023run2}. In \cite{wu2021attitude}, Wu and Sun use a control Lyapunov-barrier function (CLBF)-based controller for attitude stabilization of a rigid spacecraft with disturbances and parameter uncertainty. The work in this paper differs from that of Wu and Sun as it focuses on developing and satisfying multiple safety constraints for the attitude dynamics of the spacecraft.









The contributions of this paper are: 
\begin{enumerate}
    \item the development and application of ASIF RTA  to simultaneously enforce multiple safety constraints to a spacecraft with attitude control,
    \item the evaluation of simultaneous safety constraint satisfaction by the RTA through Monte Carlo simulation, and
    \item the analysis of a specific case study showing performance with and without RTA. 
\end{enumerate}
%
This paper is organized as follows: section~\ref{s:RTA} provides an overview of RTA, 
section~\ref{s:PF} explains the system representation, section~\ref{s:Res} discusses the results of the RTA controller, and section~\ref{s:Con} outlines conclusions and future work.

\section{Run Time Assurance} \label{s:RTA}
This work considers a continuous-time control affine system defined as,
\begin{equation}\label{e:dynamics}
    {\bm{\dot{x}}}=  f({\bm{x}})+ g({\bm{x}}){\bm{u}},
\end{equation}
where $\bm{x} \in \mathcal{X} \subseteq $ ${\mathbb{R}} ^ {n}$  denotes the state vector, ${\bm{u}} \in {\mathcal{U}} \subseteq {\mathbb{R}}^{m}$ denotes the control vector, and $f : \mathcal{X} \rightarrow  $ ${\mathbb{R}} ^ {n}$  and $g : \mathcal{X} \rightarrow $ ${\mathbb{R}} ^ {n \times m}$  are locally Lipschitz continuous functions. Here, $\mathcal{X}$ is the set of all possible state values and $\mathcal{U}$ is the admissible control set.

\begin{figure}[h!]
    \centering
    \includegraphics[width=3.25in]{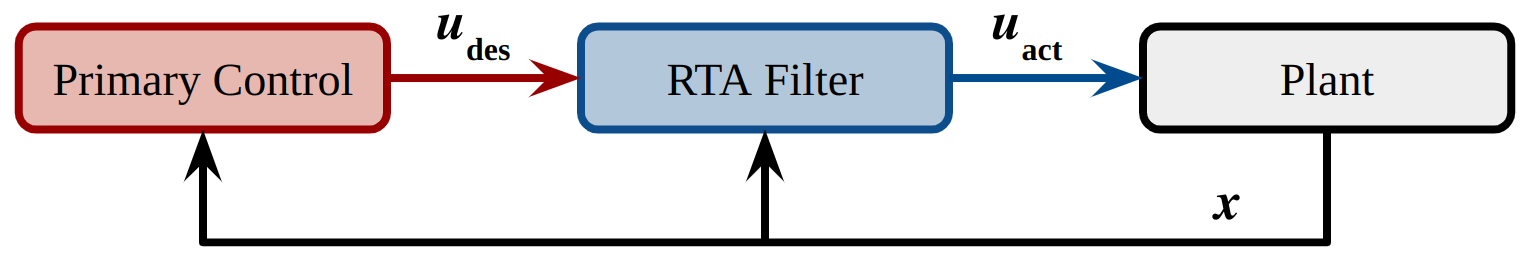}
    \caption{Control Loop with RTA.}
    \label{rta_cl}
\end{figure}

RTA is an online safety assurance technique that filters the control inputs from a primary controller to preserve safety of the system. RTA filters are decoupled from the primary controller. This enables the objective of the primary controller to be to achieve the highest performance while the objective of the RTA filter is to assure safety \cite{hobbs2023runtime}. Additionally, this permits any type of controller to be used in conjunction with RTA without impacting the safety of the system. The RTA filter intercepts the desired control $\bm{u}_{\rm des}$ from the primary controller, modifies it as necessary to assure safety, and then passes this new action $\bm{u}_{\rm act}$ to the plant. A feedback control system with RTA is shown in Figure~\ref{rta_cl}.

The ASIF RTA \cite{gurriet2018online} is an optimization-based technique designed to be minimally invasive towards the primary controller. ASIF algorithms use CBFs \cite{ames2019control} to enforce adherence to safety constraints. The ASIF algorithm used in this work is defined as,
\begin{equation}\label{e:ASIF}
    \begin{gathered}
    {\bm{u}}_{\rm act}({\bm{x}},{\bm{u}}_{\rm des}) = \argmin_{{\bm{u}} \in \mathcal{U}} \lvert \lvert {\bm{u}}_{\rm des} - {\bm{u}} \rvert \rvert ^2 \\
     \text{s.t. } BC_i({\bm{x}}, {\bm{u}}) \geq 0, \forall i\in \{1,..., M \}. 
    \end{gathered}
\end{equation}

ASIF uses a Quadratic Program (QP) to minimize the $l^2$ norm difference between ${\bm{u}}_{\rm des}$ and ${\bm{u}}_{\rm act}$, and therefore only modifies the primary control signal when necessary to assure safety. The inequality constraints for the QP are a set of $M$ barrier constraints (BC). These constraints must be defined linearly with respect to controls but can be nonlinear with respect to the states. To define safety of the system, consider a set of $M$ continuously differentiable functions $h_i : {\mathcal{X}} \rightarrow {\mathbb{R}}, \forall i \in \{1,...,M\}$, then the safe set, $\mathcal{C}_S$, can be defined as,
\begin{equation}
    {\mathcal{C}}_S := \{ {\bm{x}} \in {\mathcal{X}} \, \lvert \, h_i({\bm{x}}) \geq 0, \forall i \in \{1,...,M \}\}.
\end{equation}

The system is considered safe if $\mathcal{C}_S$ is forward invariant with respect to Eq.~\ref{e:dynamics}, such that, 
\begin{equation}
    {\bm{x}} ({t}_0) \in  {\mathcal{C}}_S \Rightarrow {\bm{x} }(t) \in{} {\mathcal{C}}_S , \forall{} \geq{} t_0.
\end{equation}

Barrier constraints, which are satisfied when $BC_i({\bm{x}},{\bm{u}}) \geq 0$, enforce Nagumo's condition \cite{nagumo1942lage}.
The boundary of the set formed by $h_i(\bm{x})$ is analyzed, where if $\dot{h}_i(\bm{x})$ is never decreasing, then $\bm{x}$ will never leave the safe set ${\mathcal{C}}_S$. Since the boundary of a set does not have volume, a strengthening class-$\kappa$ infinity function, $\alpha$, is utilized to enforce the boundary condition but relax the constraint away from the boundary. Additionally a bias, $\sigma$, can be added to the equation, incorporating a buffer on the constraint boundary to ensure it is not crossed. The resulting BC is defined as,
\begin{equation}
    BC_i({\bm{x}},{\bm{u}}) = \dot{h}_i({\bm{x}},{\bm{u}})+\alpha (h_i({\bm{x}})) + \sigma \geq 0,
\end{equation}
where $\dot{h}_i({\bm{x}})$ is defined as,
\begin{equation}
    \dot{h}_i({\bm{x}})=\nabla h_i({\bm{x}})\dot{{\bm{x}}} = L_f h_i({\bm{x}})+L_g h_i({\bm{x}}) {\bm{u}},
\end{equation}
where $L_f$ and $L_g$ are Lie derivatives of $h_i$ along $f$ and $g$ respectively. Using Eq.~\ref{e:dynamics}, the BC becomes,
\begin{equation}\label{e:BC}
    BC_i({\bm{x}},{\bm{u}}):=\nabla h_i({\bm{x}})(f({\bm{x}})+ g({\bm{x}}){\bm{u}}) + \alpha (h_i({\bm{x}}))+ \sigma.
\end{equation}

\subsection{High Order Control Barrier Functions}
For some constraints, $\nabla h({\bm{x}})$ may not depend on the control input. In this case, the BC is not valid as $\nabla h({\bm{x}})g({\bm{x}}){\bm{u}}=0$, and therefore can not be enforced by the QP. To solve this problem, High Order Control Barrier Functions (HOCBFs) can be used to differentiate the constraint and allow $\nabla h({\bm{x}})$ to depend on $\bm{u}$. A sequence of functions ${{\Psi}}_j: {\mathcal{X}} \rightarrow {\mathbb{R}}, \forall j \in \{1,...,r-1\}$ is defined as,
\begin{equation}\label{e:HOBC}
    {{\Psi}}_j({\bm{x}}):=\dot{{{\Psi}}}_{j-1}({\bm{x}}) + \alpha_j({{\Psi}}_{j-1}({\bm{x}})), \forall j \in \{1,...,r-1\},
\end{equation}
where $r$ is referred to as the relative degree of the system and is the number of times needed to differentiate the constraint in order for the control to explicitly appear in the corresponding derivative \cite{xiao2022control}. Here, $\Psi_0({\bm{x}}) = h({\bm{x}})$. In the cases where $j<r-1, \dot{\Psi}({\bm{x}}))=\nabla\Psi({\bm{x}})f({\bm{x}})$. Also note that when $r=1, \Psi_1({\bm{x}})$ would be equivalent to Eq.~\ref{e:BC}.

\subsection{Slack Variables}
When simulating a system with multiple safety constraints, there can be situations or conditions in which these constraints conflict with each other. Conflicting inequality constraints can cause the QP to fail to find a solution without violating a constraint. In order to prevent this from occurring, a slack variable $\delta$ can be added into Eq.~\ref{e:ASIF} such that it becomes,
\begin{equation}\label{e:ASIF_slack}
    \begin{gathered}
    {\bm{u}}_{\rm act}({\bm{x}},{\bm{u}}_{\rm des}) = \argmin_{{\bm{u}} \in \mathcal{U}, \delta \in \mathbb{R}} \lvert \lvert {\bm{u}}_{\rm des} - {\bm{u}} \rvert \rvert ^2 + \sum_i^M p_i\delta_i^2 \\
     \text{s.t. } BC_i({\bm{x}}, {\bm{u}}) \geq \delta_i, \forall i\in \{1,..., M \} .
    \end{gathered}
\end{equation}

In addition to $\bm{u}$, the QP also solves for the slack variables which add a buffer on the constraint. $p$ is a hyper-parameter that correlates to a scaling constant on the slack variable. Its value should be large (for example $10^{12}$) such that the slack variable is only non zero in the case that the QP cannot be solved. For the QP to have a unique solution, there can be $M-1$ slack variables, meaning at least one constraint cannot have a slack variable. Including slack variables relaxes the constraint, so they should only be included on constraints that are allowed to be violated or are low priority.


\section{Problem Formulation}\label{s:PF}
The inspection problem consists of an active ``deputy'' spacecraft that is inspecting a passive ``chief'' spacecraft in a linearized relative-motion reference frame. This section describes the dynamics of the system and the constraints that are used to enforce safety. While the motivation of this work is the inspection problem, this manuscript focuses on ensuring safety of an arbitrary primary controller via an ASIF RTA filter. As such, the simulations in the subsequent sections do not have the deputy completing inspection of the chief, rather they depict the relative motion of the deputy and the safety constraints that need to be satisfied. A primary controller that inspects the chief spacecraft is left to future work. 


\begin{figure}[h!]
    \centering
    \includegraphics[width=2.5in]{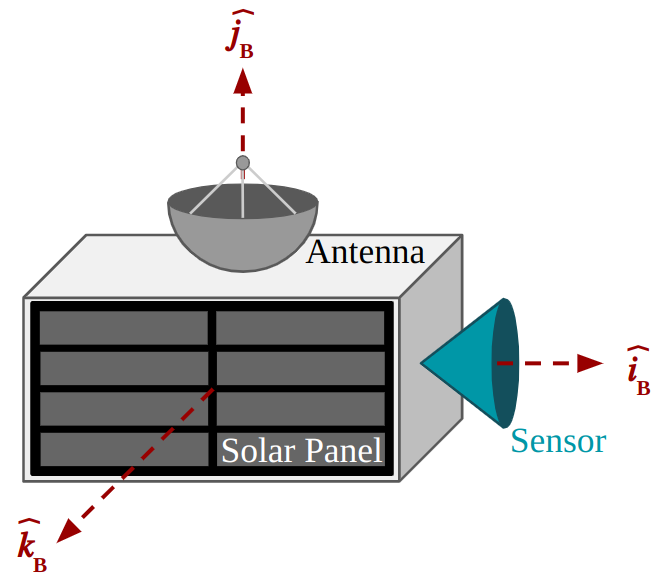}
    \caption{Hypothetical Cubesat.}
    \label{f:cube_sat}
\end{figure}

\subsection{Dynamics}
\begin{figure}[h!]
    \centering
    \includegraphics[width=1\linewidth]{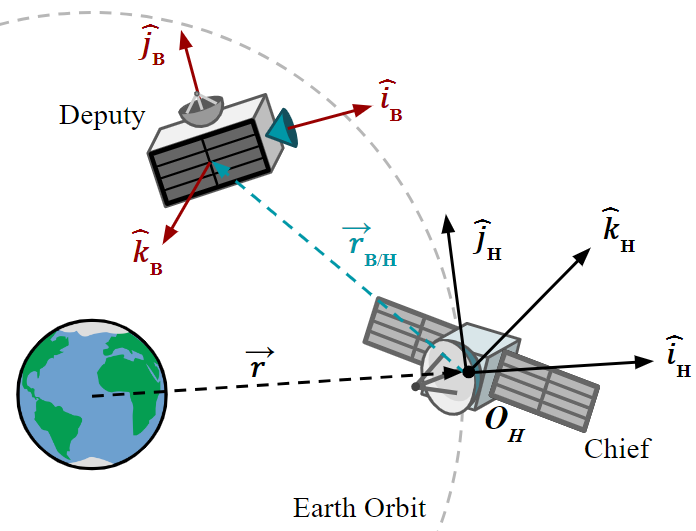}
    \caption{Hill's Reference Frame.}
    \label{f:hill_frame}
\end{figure}

The deputy spacecraft is modeled as a 6U CubeSat as shown in Figure~\ref{f:cube_sat} with the body reference frame ${\mathcal{F}}_B:=({\mathcal{O}}_B,\hat{i}_B,\hat{j}_B,\hat{k}_B)$. The spacecraft has an antenna on the $+\hat{j}_B$ face for communication with a ground station, a sensor for inspecting the chief on the $+\hat{i}_B$ face, and a solar panel on the $+\hat{k}_B$ face. The system dynamics are represented in Hill's reference frame \cite{hill1878researches}, ${\mathcal{F}}_H:=({\mathcal{O}}_H,\hat{i}_H,\hat{j}_H,\hat{k}_H)$, with the origin $O_H$ centered on the chief spacecraft in a circular orbit around Earth. As shown in Figure~\ref{f:hill_frame}, the unit vector $\hat{i}_H$ points from the center of the chief spacecraft away from the center of the Earth, in line with $\vec{r}$, $\hat{j}_H$ points in the direction of motion of the chief spacecraft around the Earth, and $\hat{k}_H$ is normal to $\hat{i}_H$ and $\hat{j}_H$. The location of the deputy relative to the chief is shown by $\vec{r}_{B/H}$. For this problem, the chief spacecraft is assumed to remain in the same circular orbit in Low Earth Orbit (LEO). The deputy spacecraft maneuvers relative to the chief spacecraft with orientation control via a three wheel reaction wheel array aligned with the principal axis. 

This subsection outlines the dynamical models used for the motion, surface temperature, and stored energy of the spacecraft as well as the dynamics for the sun in the simulation.

\subsubsection{Attitude}
The attitude of the deputy spacecraft is modeled in the principal axis body reference frame, where the origin, ${\mathcal{O}}_B$ is located at the center of mass of the spacecraft, and the axes are aligned with the eigenvectors of the spacecraft inertia matrix. The attitude is then modeled using a quaternion formulation \cite{markley2014fundamentals}, which represents a rotation between two reference frames. In this case, the quaternion is used to represent the rotation of the spacecraft from Hill's reference frame to the body reference frame. The quaternion also avoids singularity points that arise from Euler angles. A quaternion ${\mathbf{q}} \in {\mathbb{R}}^4$ is comprised of a three vector part ${\mathbf{q}}_v= {\mathbf{q}}_{1:3}$ and a scalar part $q_4$. The angular rates ${\bm{\omega}} \in {\mathbb{R}}^3$ are then used to represent the rotation of the spacecraft about each of its principal axes. The derivative of the angular rates is given by Euler's rotation equations,
\begin{equation}\label{e:eulerRot}
    {\mathbf{J}\dot{\bm{\omega}}+\bm{\omega} \times \bf{J}\bm{\omega} = \bm{\tau}},
\end{equation}
where $\mathbf{J}$ is the spacecraft's inertia matrix and $\bm{\tau}$ is the torque applied to the spacecraft via the reaction wheels. Therefore, ${\bm{\tau}} = D{\dot{\bm{\psi}}}$, where $D$ is the reaction wheel spin axis mass moment of inertia and ${\dot{\bm{\psi}}}$ is the reaction wheel acceleration. Eq.~\ref{e:eulerRot} can then be written in the principal axis body reference frame as,
\begin{equation}
    \begin{bmatrix} 
        \dot{\omega}_1 \\ \dot{\omega}_2 \\ \dot{\omega}_3
    \end{bmatrix} =
    \begin{bmatrix}
        J_1^{-1}((J_2-J_3)\omega_1\omega_3+D\dot{\psi}_1) \\ 
        J_2^{-1}((J_3-J_1)\omega_2\omega_3+D\dot{\psi}_2) \\ 
        J_3^{-1}((J_1-J_2)\omega_1\omega_2+D\dot{\psi}_3)
    \end{bmatrix}.
\end{equation}
Here, $J_1,J_2,$ and $J_3$ are the principal moments of inertia of the spacecraft. The relation of the quaternion derivatives to the angular rates is given by,
\begin{equation}
    {\mathbf{\dot{q}}} = \frac{1}{2}{\mathbf{\Xi}}(\mathbf{q})\bm{\omega},
\end{equation}
where $\mathbf{\Xi}$ is given by,
\begin{equation}
    {\mathbf{\Xi}}({\mathbf{q}})=\begin{bmatrix}
        q_s{\mathcal{I}}-{\mathbf{q}}_v^\chi \\ -{\mathbf{q}}_v^T
    \end{bmatrix} = \begin{bmatrix}
        q_4 & -q_3 & q_2 \\ q_3 & q_4 & -q_1 \\ -q_2 & q_1 & q_4 \\ -q_1 & -q_2 & -q_3
    \end{bmatrix},
\end{equation}
where ${\mathcal{I}} \in {\mathbb{R}}^3$ represents the identity matrix and $^\chi$ indicates the skew-symmetric matrix. The state vector for the attitude of the spacecraft is ${\bm{x}} = [{\mathbf{q},\bm{\omega},\bm{\psi}}]^T$, the control vector is ${\mathbf{u=\dot{\bm{\psi}}}} \in \mathbf{\mathcal{U}}$, and the dynamics are given by,
\begin{equation}
    \begin{bmatrix}
        \dot{q}_1 \\ \dot{q}_2 \\ \dot{q}_3 \\ \dot{q}_4 \\ \dot{\omega}_1 \\ \dot{\omega}_2 \\ \dot{\omega}_3 \\ \dot{\psi}_1 \\ \dot{\psi}_2 \\ \dot{\psi}_3 
    \end{bmatrix} = \begin{bmatrix}
        \frac{1}{2}(q_4\omega_1-q_3\omega_2+q_2\omega_3) \\
        \frac{1}{2}(q_3\omega_1+q_4\omega_2-q_1\omega_3) \\
        \frac{1}{2}(-q_2\omega_1+q_1\omega_2+q_4\omega_3) \\
        \frac{1}{2}(-q_1\omega_1-q_2\omega_2-q_3\omega_3) \\
        J_1^{-1}((J_2-J_3)\omega_1\omega_3+D\dot{\psi}_1) \\ J_2^{-1}((J_3-J_1)\omega_2\omega_3+D\dot{\psi}_2) \\ J_3^{-1}((J_1-J_2)\omega_1\omega_2+D\dot{\psi}_3)\\ u_1 \\ u_2 \\ u_3        
    \end{bmatrix}.
\end{equation}
The values used for this work are $D=4.1\times10^-5\:kg-m^2, J_1 = 0.022\:kg-m^2, J_2 = 0.044\:kg-m^2,$ and $J_3 = 0.056\:kg-m^2$ \cite{petersen2021challenge}.




\subsubsection{Temperature}
\begin{figure}[h!]
    \centering
    \includegraphics[width=2in]{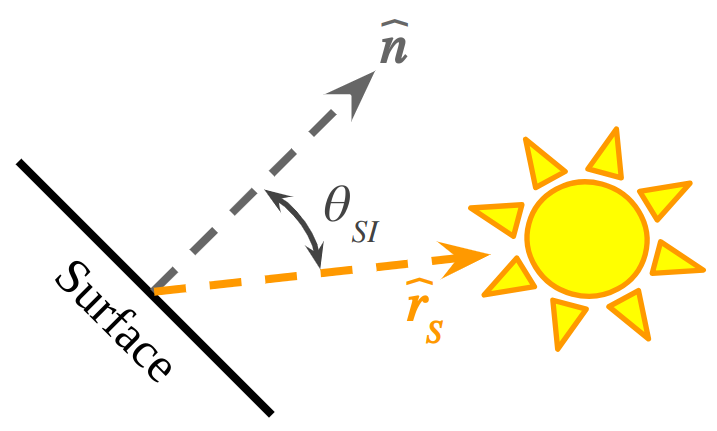}
    \caption{Sun Incidence Angle.}
    \label{f:sun_in}
\end{figure}

The temperature of the spacecraft is modeled using thermal nodes, where each side of the spacecraft is represented by one node. For this analysis, it is assumed that each node is independent, and heat is not transferred between nodes. To determine the temperature of each node, four environmental heat fluxes are considered: solar flux, Earth albedo radiation, Earth Infared Radiation (IR), and heat rejection from the spacecraft \cite{foster2022small}. Solar flux, which is radiant energy from the sun, is the largest heat source for the spacecraft in LEO. The heat transfer, $\dot{q}$, from solar flux is given by,
\begin{equation} \label{e:t_sol}
    \dot{q}_{solar}=\alpha{}AS(\hat{n}\cdot \hat{r}_S),
\end{equation}
where $\alpha$ is the absorptivity of the surface, $A$ is the surface area, $S=1367\:W/m^2$ is the solar constant \cite{wertz1999space}, $\hat{n}$ is the surface normal vector, and $\hat{r}_S$ is the unit vector pointing from the surface to the sun. The sun incidence angle $\theta_{SI}$ is formed from $\hat{n}$ and $\hat{r}_S$ as shown in Figure~\ref{f:sun_in}.

Note that $\theta_{SI}$ is only defined on the interval $[0,\frac{\pi}{2}]$, and therefore $\dot{q}_{solar}=0$ when $\hat{n}\cdot \hat{r}_S \leq 0$.
Earth albedo radiation is the solar energy that is reflected by the Earth, and is dependent on the conditions of the surface of the Earth. The transfer is given by, 
\begin{equation}\label{e:t_albedo}
    \dot{q}_{albedo}=\alpha A S A_f F_E
\end{equation}
where $A_f$ is the albedo factor of Earth, which is assumed to be $0.27$ \cite{wertz1999space}, and $F_E$ is the view factor between the surface and earth. Determination of the view factor for Earth can become complex \cite{garzon2018thermal}, but for this analysis it is simplified to $F_E=0.8\cos{\theta_{EI}}$ , where $\theta_{EI}$ is the incidence angle between the surface normal vector and the unit vector pointing from the surface to the Earth $\hat{r}_E$. In Hill's reference frame, the Earth always points along the $-\hat{x}$ axis, therefore $\hat{r}_E=[-1,0,0]^T$.

Earth IR is heat emitted by the Earth, where the heat transfer is given by,
\begin{equation}\label{e:t_ir}
    \dot{q}_{IR}=\sigma \epsilon A F_E T_E^4
\end{equation}
where $\sigma=5.67051\times10^{-8}\:W/(m^2\times K^4)$ is the Stefan-Boltzmann constant \cite{wertz1999space}, $\epsilon$ is the surface emissivity, and $T_E$ is the temperature of the Earth, which on average is $255\:K$ \cite{foster2022small}.

Heat is rejected by the spacecraft into open space through radiation, where the heat transfer is given by,
\begin{equation}\label{e:t_rej}
    \dot{q}_{rejected} = \sigma \epsilon A T^4
\end{equation}
where $T$ is the temperature of the surface and the temperature of space is assumed to be zero. 
The total heat transfer for a node is then given by,
\begin{equation}
    \dot{q}_{total} = \dot{q}_{solar}+\dot{q}_{albedo}+\dot{q}_{IR}-\dot{q}_{rejected}
\end{equation}
The derivative of the temperature is then given by
\begin{equation}
    \dot{T}=\frac{\dot{q}_{total}}{mc_p}
\end{equation}
where $m$ is the mass of the node and $c_p$ is the specific heat of the material. For this paper, the temperature of one side of the spacecraft is tracked, which has $m=2\:kg$ and $A=300\:cm^2$. This side is assumed to be made of aluminum, which has the properties: $c_p=900\:J/(kg\times K)$, $\alpha=0.13$, and $\epsilon=0.06$ \cite{wertz1999space}.

\subsubsection{Energy}
For this simulation, the spacecraft has a battery and solar panels to charge the battery. In order to generate power, the solar panels must be facing the sun, where $\theta_{SI}\in [-\frac{\pi}{2},\frac{\pi}{2}]$. The power generated by the solar panels, $P_{in}$, is given by,
\begin{equation}
    P_{in} = P_I I_d A \cos{\theta_{SI}}
\end{equation}
where $P_I$ is the ideal performance and $I_d$ is the inherent degradation of the solar panels. The change in energy of the battery is given by,
\begin{equation}
    \dot{E}=P_{in}-P_{out}
\end{equation}
where the energy used by the spacecraft, $P_{out}$, is assumed to be a constant value of $15\:W$. The solar panels characteristics are: $P_I =983.3\:W$ \cite{bluePower} and $I_d=0.77$ \cite{wertz1999space}.

\subsubsection{Sun}
With the Earth and spacecraft fixed in Hill's reference frame, the sun is considered to rotate around the spacecraft. For this analysis, it is assumed that the sun rotates at a constant rate in the $\hat{x}-\hat{y}$ plane. The unit vector pointing from the center of the spacecraft to the sun, $\hat{r}_S$, is defined as,
\begin{equation}
    \hat{r}_S = [\cos{\theta_S},\cos{\theta_S},0].
\end{equation}

The change in angle $\theta_S$ is dependent on the spacecraft mean motion, $n$, such that $\dot{\theta}_S=-n=-0.001027\:rad/s$. In terms of angular rates, this is given by ${\bf{\omega}} = [0,0,-0.001027]^T$. Note, while the Sun is considered to move while the earth is stationary, it is assumed that the Earth never blocks sunlight from reaching the spacecraft.

\subsubsection{Total System Dynamics}
Combining the dynamics formulated in the previous subsections, the full state vector of the system becomes ${\bm{x}} = [{\bf{q}},{\bm{\omega}},{\bm{\psi}},T,E,\theta_S,x,y,z,\dot{x},\dot{y},\dot{z}]^T$, and the control vector for the system is ${\bm{u}} = [{\bm{\dot{\psi}}}]$. For the ASIF RTA, the functions $f({\bm{x}})$ and $g({\bm{x}})$ are populated using the full state vectors and the dynamics detailed in previous subsections.

\subsection{Constraints}\label{s:constraints}
This section details the constraints on the spacecraft attitude. The constraints are derived from rotational safety requirements that were generated from concerns of satellite operators and formally described by Hobbs, et al. in \cite{hobbs2021risk}. The constraints are formatted as inequality constraints $h({\bm{x}})$ but implemented as BCs as in Eq.~\ref{e:BC} or Eq.~\ref{e:HOBC}.


\subsubsection{Attitude Exclusion Zone}
\begin{figure}[h!]
    \centering
    \includegraphics[width=2.5in]{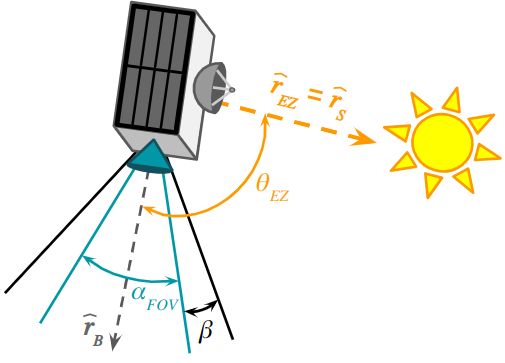}
    \caption{Attitude Exclusion Zone.}
    \label{f:koz}
\end{figure}

The spacecraft shall adhere to conical attitude exclusion zones. For example, to avoid instrument blinding, it may be necessary to prevent a sensor from pointing directly towards the Sun. Consider the unit vectors pointing along the sensor boresight, $\hat{r}_B$, and from the center of the spacecraft to the exclusion zone, $\hat{r}_{EZ}$. The angle between these two unit vectors, $\theta_{EZ}$, is found using the dot product as shown in Figure~\ref{f:koz}. 

The attitude exclusion zone is then defined as,
\begin{equation}
    h_{EZ}({\bm{x}}) = \theta_{EZ} - \frac{\alpha_{FOV}}{2}-\beta,
\end{equation}
where $\alpha_{FOV}$ is the sensor's field of view and $\beta$ is a safety buffer. This constraint is relative degree 2. Therefore, HOCBFs are used to transform $h_{EZ}$ into a valid CBF with Eq.~\ref{e:HOBC}. For this simulation, $\hat{r}_{EZ}=\hat{r}_S$, $\hat{r}_{B}=\hat{n}_x$, $\alpha_{FOV}=60\:deg$, and $\beta = 10\:deg$. 

\subsubsection{Communication}
\begin{figure}[h!]
    \centering
    \includegraphics[width=2.5in]{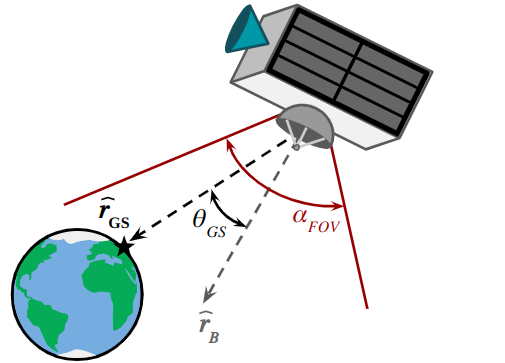}
    \caption{Communication Angle.}
    \label{f:kiz}
\end{figure}

The spacecraft shall maintain attitude requirements for communication with the ground station. Consider the unit vectors pointing along the transmitters boresight, $\hat{r}_B$, and from the center of the spacecraft to the ground station, $\hat{r}_{GS}$. The angle between these two unit vectors, $\theta_{GS}$, is found using the dot production, and is shown in Figure~\ref{f:kiz}. 

The communication constraint is then defined as,
\begin{equation}
    h_{comm}({\bm{x}}) = \frac{\alpha_{FOV}}{2}-\theta_{GS}.
\end{equation}

This constraint is relative degree 2; therefore, HOCBFs are used to transform $h_{comm}$ into a valid CBF with Eq.~\ref{e:HOBC}. For this simulation, $\hat{r}_{EZ}=\hat{r}_S$, $\hat{r}_{B}=\hat{n}_x$, $\alpha_{FOV}=60\: deg$, and $\beta = 10\:deg$. For this simulation, $\hat{r}_{GS}=\hat{r}_EI$, $\hat{r}_{B}=\hat{n}_y$, and $\alpha_{FOV}=180\: deg$.

\subsubsection{Maintain Acceptable Temperature}
The spacecraft shall maintain acceptable temperature of all components. For example, certain components of the spacecraft may overheat, in which case the spacecraft shall maneuver to rotate the component away from significant heat sources, typically the Sun. Given the temperature of a component, $T$, the temperature constraint is defined simply as,
\begin{equation}
    h_{temp}(\bm{x}) = T_{max}-T.
\end{equation}

This constraint is relative degree 3 and could be enforced using HOCBFs; however, the Sun and Earth incidence angles with the surface are added to achieve better performance. The constraint is then given as,
\begin{equation}
    \Psi_{temp}({\bm{x}}) = T_{max}-T-\delta_0(\frac{\pi}{2}-\theta_{SI})-\delta_1(\frac{\pi}{2}-\theta_{EI})
\end{equation}
where $\delta_0$ and $\delta_1$ are constant coefficients that are tuned external to the QP and $T_{max}$ is the maximum allowable temperature. This constraint is relative degree 2, where HOCBFs are then used to transform $\Psi_{temp}$ into a valid CBF. Note that when $\theta_{SI} > 90\:deg$ and $\theta_{EI}>90 \:deg$, the gradient of the constraint is zero. However, in this case, temperature is decreasing and is not approaching $T_{max}$, so the CBF is valid. For this simulation, $T_{max}=10\: ^o C$ and $T$ is measured on the side of the spacecraft pointing along the $-\hat{j}_B$ axis.

\subsubsection{Maintain Battery Charge}
The spacecraft shall maintain attitude requirements for sufficient power generation. For example, the spacecraft shall point its solar panels towards the sun to ensure the battery remains charged. Given the battery energy, $E$, the battery charge constraint is defined simply as,
\begin{equation}
    h_{batt}({\bm{x}}) = E-E_{min}
\end{equation}
where $E_{min}$ is the minimum allowable energy level, determined such that at this point the spacecraft would be able to maintain minimum function as determined by the mission. This constraint is relative degree 3. In this case, note that the gradient of the constraint is zero when the sun incidence angle with the side of spacecraft with the solar panels, $\theta_{SI}$, is greater than $90\:deg$. Therefore, HOCBFs cannot be used to enforce this constraint. To account for this, $\theta_{SI}$ is added to the constraint, which is then given by,
\begin{equation}
    \Psi({\bm{x}}) = E-E_{min}-\delta_2\theta_{SI}.
\end{equation}

This constraint is relative degree 2, where HOCBFs are then used to transform it into a valid CBF. For this simulation, $E_{min}=1\:kJ$, and solar panels are assumed to cover the side of the spacecraft pointing along the $+z_B$ axis.

\subsubsection{Avoid Actuation Saturation}
The spacecraft actuation strategy shall conserve actuator use to prevent wear when possible. This constraint is not enforced with RTA, but rather is implemented as the admissible control set ${\mathcal{U}}$. For this simulation, $\dot{\psi}_{max}=181.3\:rad/s^2$ and ${\mathcal{U}}=[-\dot{\psi}_{max},\dot{\psi}_{max}]^3$.

\subsubsection{Avoid Aggressive Maneuvering}
The spacecraft shall maneuver below a threshold to prevent structural damage. For example, if the spacecraft rotation accelerates too quickly components might break. Additionally, if the angular velocity is too high, it limits how responsive the spacecraft can be to new pointing commands. The avoid aggressive maneuvering constraint enforces limits on the angular velocity, $\bm{\omega}$, and angular acceleration, $\dot{\bm{\omega}}$, as well as the reaction wheel velocity, $\bm{\psi}$. The constraints for angular velocity and reaction wheel velocity are defined as,
\begin{equation}
    h_{\omega}({\bm{x}})=\omega_{max}^2-\omega^2
\end{equation}
\begin{equation}
    h_{\psi}({\bm{x}})=\psi_{max}^2-\psi^2
\end{equation}

These constraints are relative degree 1 and are valid CBFs. Since the angular acceleration is not a state, it must be enforced differently. Given the system dynamics, $\dot{\omega}$ is directly related to $\dot{\psi}$ and $\omega$, so additional bounds for $\dot{\psi}_{max}$ can be used to enforce the constraint. The maximum angular acceleration occurs when ${\bm{\omega}} = [\omega_{max}]^3$ and  ${\dot{\bm{\omega}}} = [\dot{\omega}_{max}]^3$. From the full system dynamics, enforcing the following constraints ensures that $|\dot{\omega}|\leq \dot{\omega}_{max}$,
\begin{equation}
\begin{split}
    |\dot{\psi}_1| \leq \frac{J_1\dot{\omega}_{max} - |J_2-J_3|\omega_{max}^2}{D} \\
    |\dot{\psi}_2| \leq \frac{J_2\dot{\omega}_{max} - |J_3-J_1|\omega_{max}^2}{D} \\
    |\dot{\psi}_3| \leq \frac{J_3\dot{\omega}_{max} - |J_1-J_2|\omega_{max}^2}{D}.
    \end{split}
\end{equation}

These constraints are enforced as additional bounds on the admissible control set ${\mathcal{U}}$. For this simulation, $\dot{\omega}_{max} = 2\:deg/s^2$, $\omega_{max}=1\:deg/s$, and $\psi_{max}=576\:rad/s$.

\section{Simulation Results} \label{s:Res}
This section presents the results of a simulation developed to test the constraints outlined in Section~\ref{s:constraints}, where ASIF RTA is used to assure safety. This simulation uses the Safe Autonomy Run Time Assurance Framework \cite{ravaioli2023universal}, which simplifies user effort and supports automatic differentiation of all constraints, becoming particularly useful for the constraints that utilize HOCBFs. In these simulations, if the QP fails, the desired control (${\bm{u}}_{\rm des}$) is passed through to the spacecraft. The simulation is without disturbances, and states measurements are exact, without noise or errors. As detailed in Section~\ref{s:RTA}, slack variables can be utilized when a system has multiple constraints that can be conflicting. In this work, there are a total of six separate constraints. From initial testing, it was found that the communication constraint was the most restrictive and the one that was violated most often. Due to satellites having specific communication windows with ground stations there is not a need for the spacecraft to always point at the earth. As such, a slack variable was added to this constraint.

\subsection{Active Primary Controller Simulation}

\begin{figure*}[h!]
    \centering
    \subfigure[]{\includegraphics[width=0.3\textwidth]{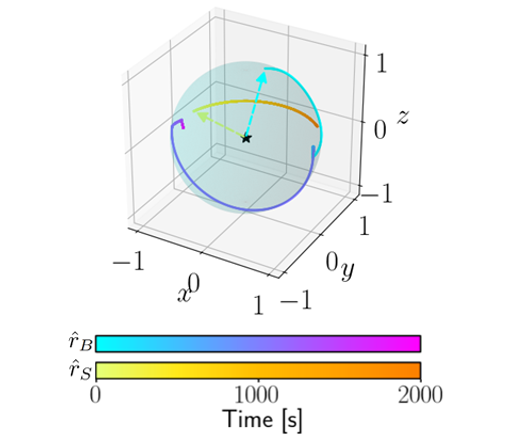}\label{f:f:cvec}}
    \subfigure[]{\includegraphics[width=0.3\textwidth]{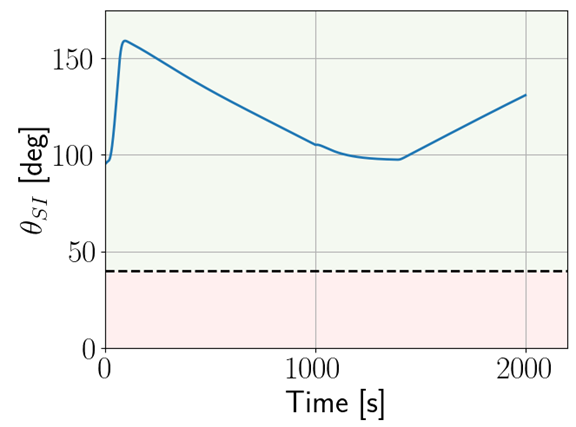}\label{f:f:csun}}
    \subfigure[]{\includegraphics[width=0.3\textwidth]{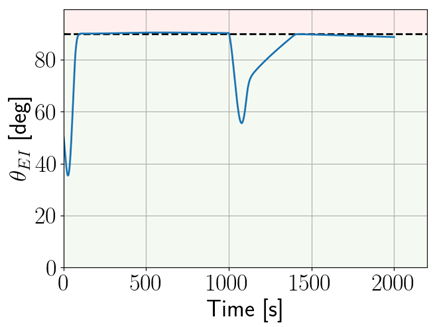}\label{f:f:cearth}}
    
    \subfigure[]{\includegraphics[width=0.3\textwidth]{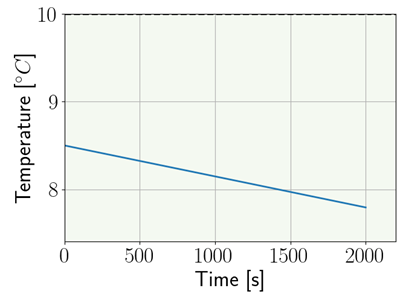}\label{f:f:ctemp}}
    \subfigure[]{\includegraphics[width=0.3\textwidth]{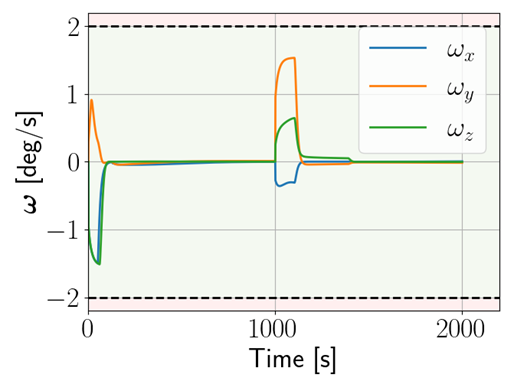}\label{f:f:cw}}
    \subfigure[]{\includegraphics[width=0.3\textwidth]{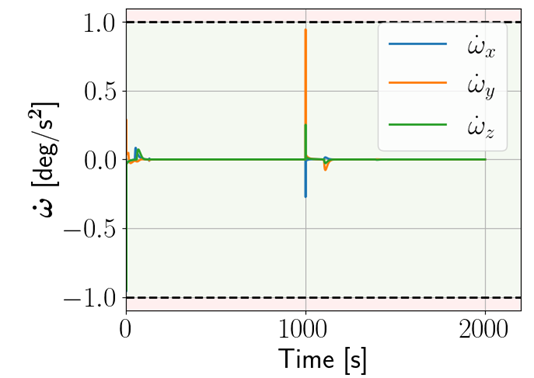}\label{f:f:cwdot}}
    
    \subfigure[]{\includegraphics[width=0.3\textwidth]{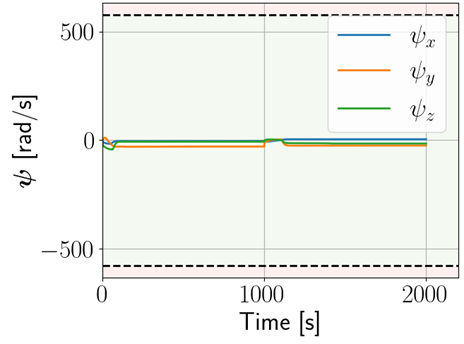}\label{f:f:cpsi}}
    \subfigure[]{\includegraphics[width=0.3\textwidth]{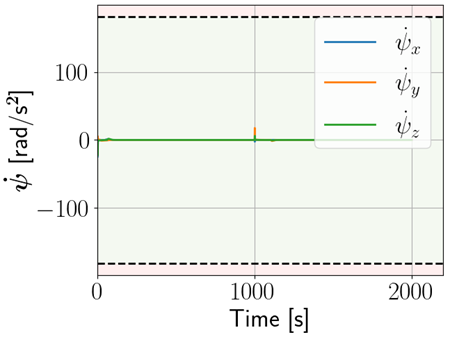}\label{f:f:cpsidot}}
    \subfigure[]{\includegraphics[width=0.3\textwidth]{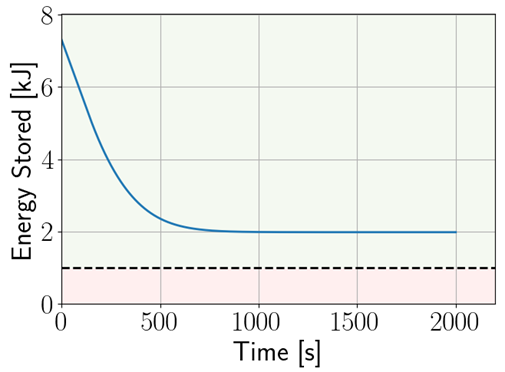}\label{f:f:cenergy}}
    \caption{Example of a simulation with RTA and primary controller active.}
    \label{f:controlledEx}
    
\end{figure*}

A primary controller was implemented into the simulation to better demonstrate ASIF RTA intervening. The primary controller for this simulation is a simple PD controller that attempts to track a commanded quaternion $\bm{{q}}_c$ \cite{markley2014fundamentals}. The primary controller is designed to be aggressive and violate the constraints in order to show that RTA is able to assure safety. The primary control law is given by,
\begin{equation}
    {\bm{u}}_{des} = {\dot{\bm{\psi}}}_{max} \tanh(-k_p{\delta\bm{q}}_{1:3}-k_d{\bm{\omega}}),
\end{equation}
where $k_p$ is the proportional gain, $k_d$ is the derivative gain, and,
\begin{equation}
\begin{split}
    {\bm{\delta q}}_{1:3}=q_{c4}^{-1}{\bm{q}}_{1:3}+q_4{\bm{q}}_{c1:3}^{-1}-{\bm{q}}_{1:3}\times {\bm{q}}_{c1:3}^{-1}\\
    = [-q_{c1}, -q_{c2}, -q_{c3}, -q_{c4}]/\lvert \lvert{\bm{{q}}_c} \rvert \rvert^2.
    \end{split}
\end{equation}
The values $k_p=0.2$ and $k_d=1.5$ are used, and ${\dot{\bm{\psi}}}_{max}$ are used to ensure ${\bm{u}}_{des} \in {\mathcal{U}}$. The simulation is run for 2,000 seconds with a time interval of 1 second. For the first 1,000 seconds, ${\bm{{q}}_c}=[0,0,0,1]$, and for the last 1,000 seconds, ${\bm{{q}}_c}=[0,1,0,0]$. The initial conditions for this simulation are ${\bm{q}}=[0.680, -0.151, 0.630, 0.343], {\bm{\omega}}=[0,0,0] \,deg/s$, ${\bm{\psi}}=[0,0,0] \, rad/s$, $T=8.5^o C$, $E=7.3 \,kJ$, and $\theta_{S} = 525^o$. Results of the simulation are shown in Figure~\ref{f:controlledEx} and show that the ASIF RTA is able to assure safety over all constraints for the duration of the simulation, while the communication constraint is minimally violated due to the slack variable.  In the same manner as Figure~\ref{f:mc}, the boundary of the safe set is denoted by black dashed lines, the green shaded region is safe, and the red shaded region is in violation of the constraint. Figure~\ref{f:f:cvec} shows the trajectories of the sun vector $\bm{\hat{r}}_S$ and the sensor bore-sight $\bm{\hat{r}}_B$ over time, where $\bm{\hat{r}}_S$ changes from yellow to dark orange and $\bm{\hat{r}}_B$ changes from light blue to purple as time increases. Figure~\ref{f:f:csun} depicts the angle between the bore-sight of the sensor and the sun. Figure~\ref{f:f:cearth} depicts the angle between the communication antenna and the earth.

\begin{figure*}[h!]
    \centering
    \subfigure[]{\includegraphics[width=0.3\textwidth]{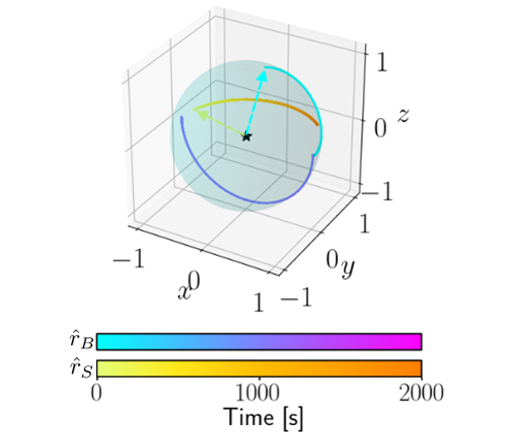}}
    \subfigure[]{\includegraphics[width=0.3\textwidth]{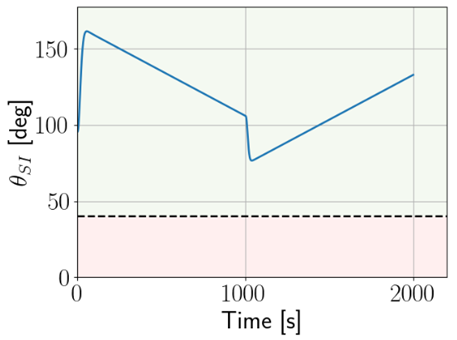}}
    \subfigure[]{\includegraphics[width=0.3\textwidth]{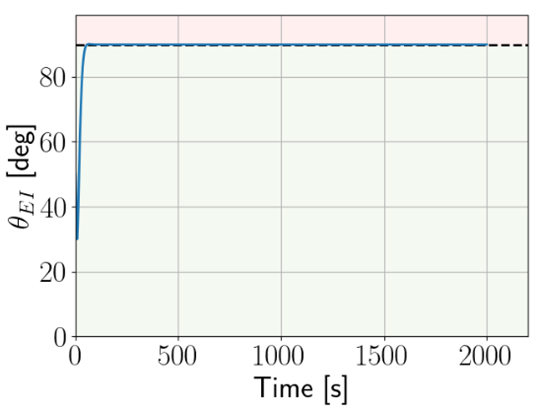}}
    
    \subfigure[]{\includegraphics[width=0.3\textwidth]{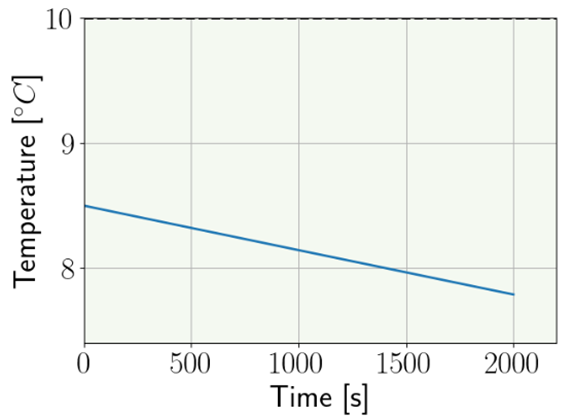}}
    \subfigure[]{\includegraphics[width=0.3\textwidth]{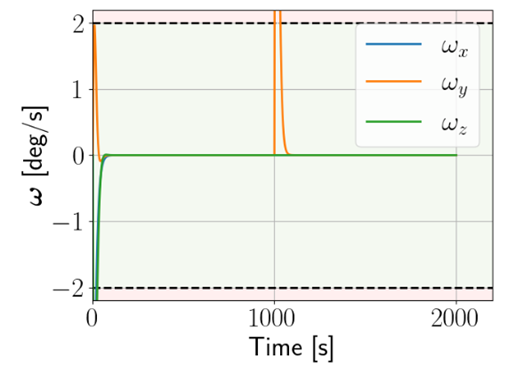}}
    \subfigure[]{\includegraphics[width=0.3\textwidth]{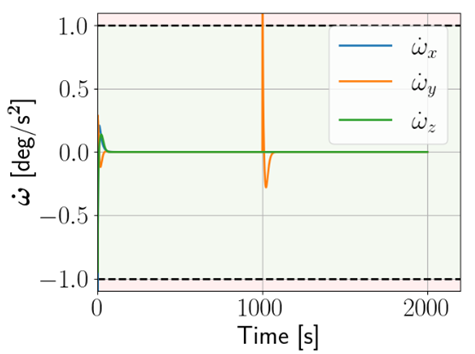}}
    
    \subfigure[]{\includegraphics[width=0.3\textwidth]{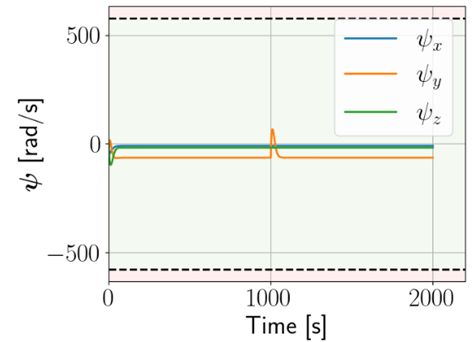}}
    \subfigure[]{\includegraphics[width=0.3\textwidth]{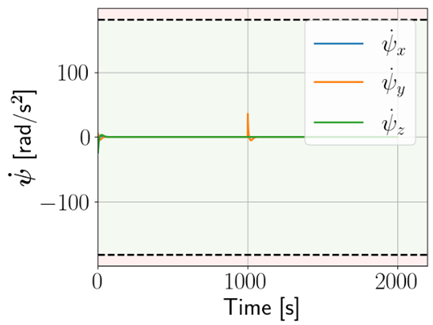}}
    \subfigure[]{\includegraphics[width=0.3\textwidth]{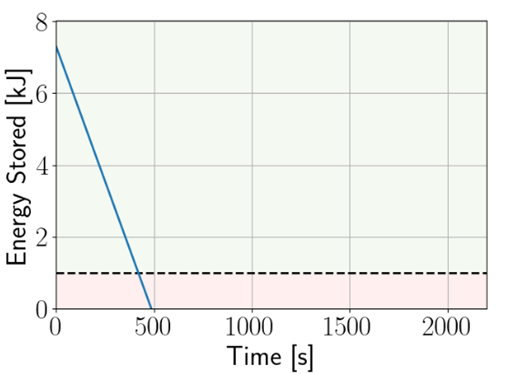}}
    \caption{Example of a simulation with primary controller active without RTA that violates angular dynamics and energy storage constraints.}
    \label{f:controlledEx_noRTA}
\end{figure*}

Figure~\ref{f:controlledEx_noRTA} has the same initial conditions and active primary controller as Figure~\ref{f:controlledEx} but does not have RTA implemented. As can be seen from the plots, this simulation does not stay within the defined constraints of the problem. Specifically y-axis angular velocity and acceleration well exceed the limit values within seconds of the commanded quaternions. The stored energy decrease from the start of the simulation, crossing the threshold limit of $1 \, kJ$ around 400 seconds into the simulation and never increases to an acceptable level throughout the remainder of the simulation. This simulation demonstrates that the specified initial conditions and active primary controller cause the system to violate the safety constraints.

\subsection{Monte-Carlo Analysis}
To verify that the RTA assures safety regardless of initial conditions, a Monte-Carlo experiment was conducted. The simulations begin with random initial conditions that are within the safe set, meaning no constraints are violated. To prevent a state being initialized on the constraint boundary, a $5\%$ buffer is added. For this analysis, 5,000 simulations were run for 2,000 seconds each with a time interval of 1 second. The spacecraft did not have any primary control input, and Latin hypercube sampling was used to generate initial conditions. \begin{figure*}[h!]
    \centering
    \subfigure[]{\includegraphics[width=0.3\textwidth]{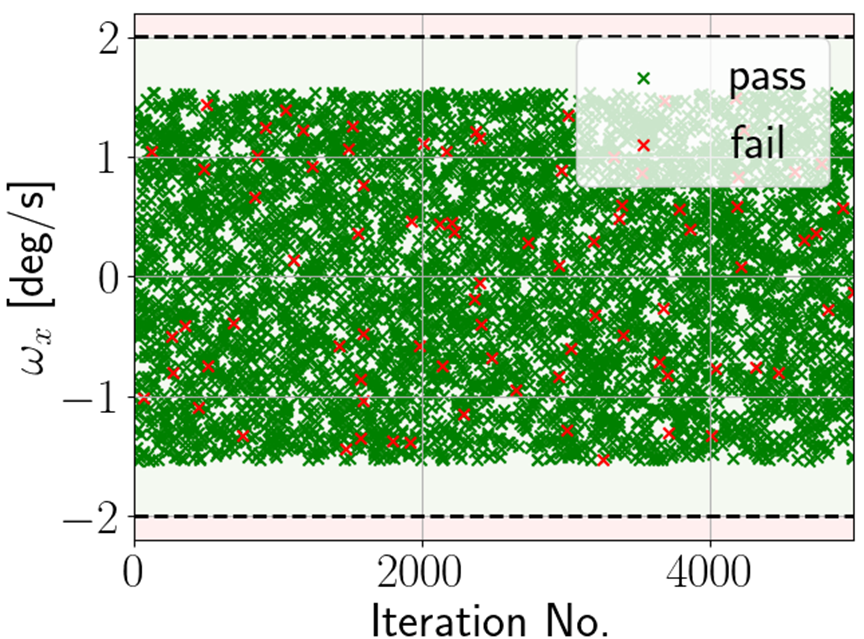}\label{sf:ic:w1}}
    \subfigure[]{\includegraphics[width=0.3\textwidth]{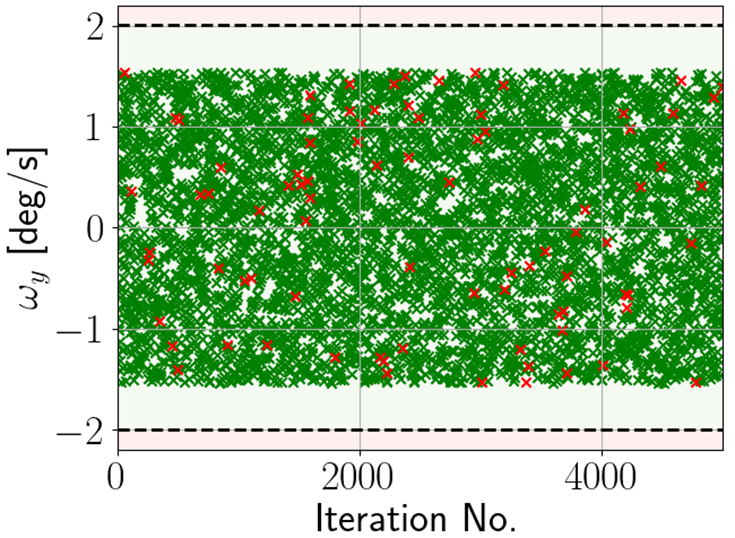}\label{sf:ic:w2}}
    \subfigure[]{\includegraphics[width=0.3\textwidth]{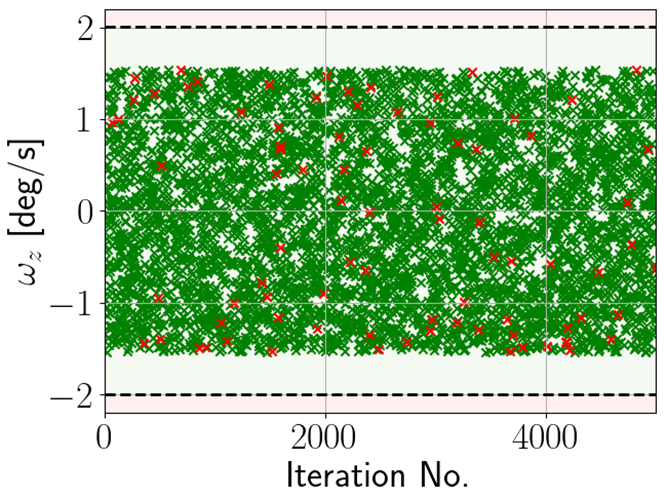}\label{sf:ic:w3}}
    
    \subfigure[]{\includegraphics[width=0.3\textwidth]{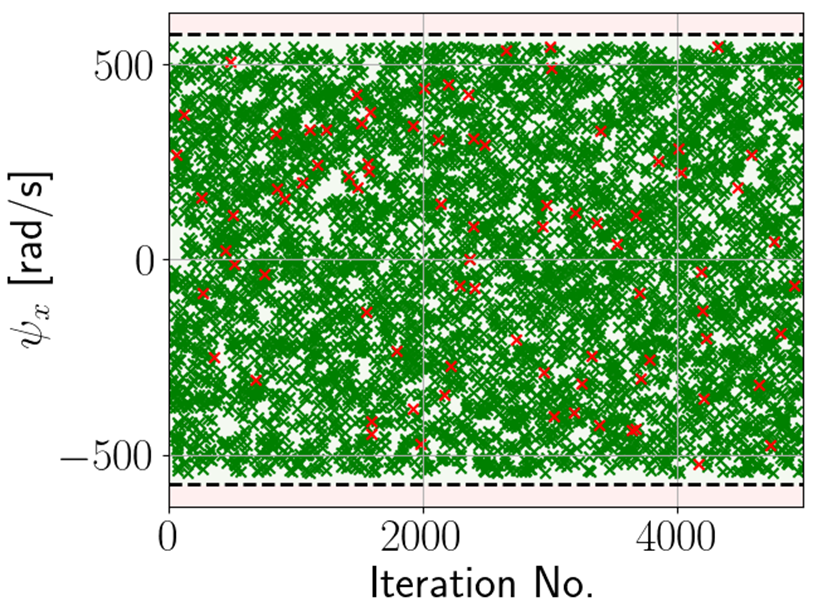}\label{sf:ic:psi1}}
    \subfigure[]{\includegraphics[width=0.3\textwidth]{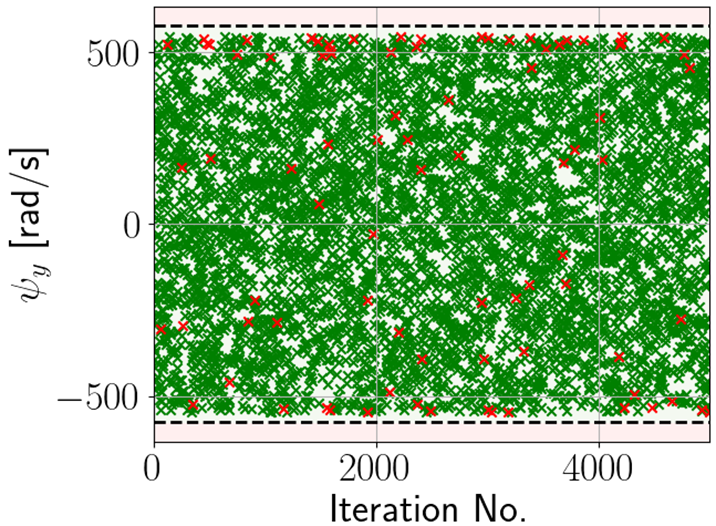}\label{sf:ic:psi2}}
    \subfigure[]{\includegraphics[width=0.3\textwidth]{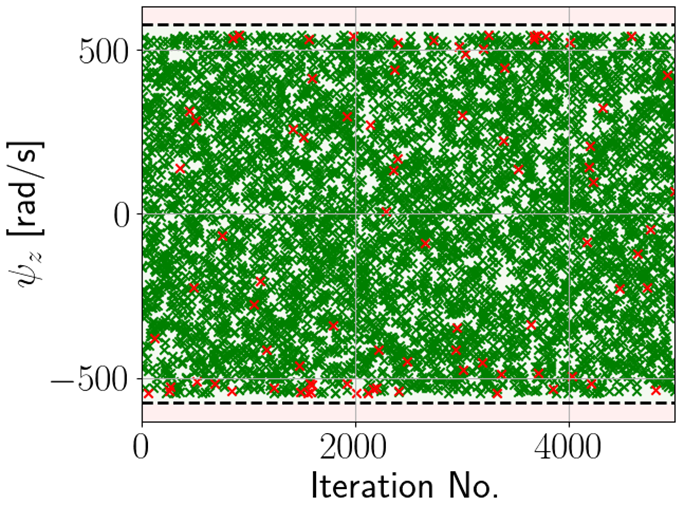}\label{sf:ic:psi3}}
    
    \subfigure[]{\includegraphics[width=0.3\textwidth]{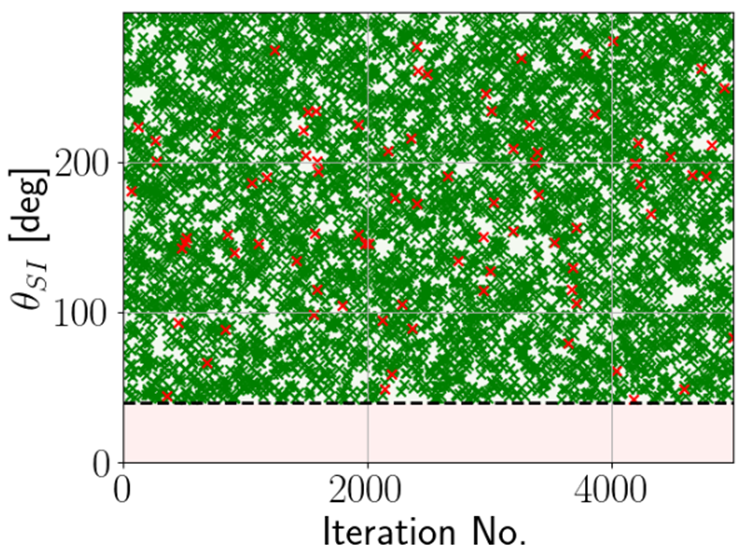}\label{sf:ic:tS}}
    \subfigure[]{\includegraphics[width=0.3\textwidth]{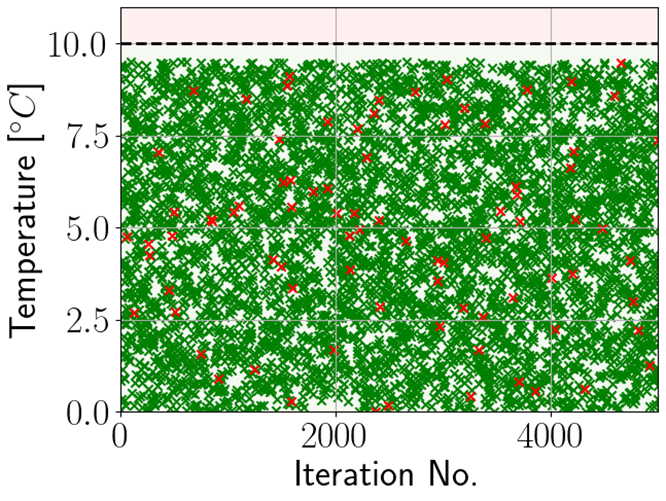}\label{sf:ic:temp}}
    \subfigure[]{\includegraphics[width=0.3\textwidth]{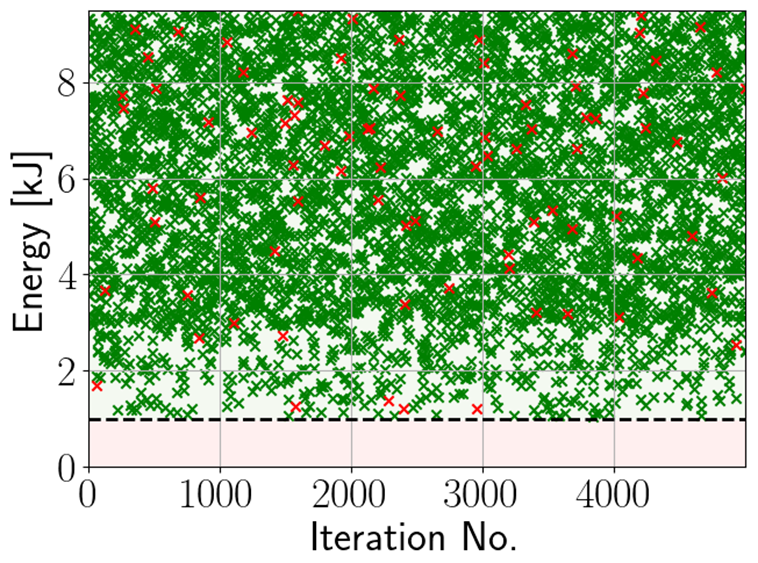}\label{sf:ic:en}}
    
    \caption{Initial condition test points from Monte-Carlo Analysis.}
    \label{f:mc}
\end{figure*}

Figure~\ref{f:mc} shows the results of the Monte-Carlo analysis. The boundary of the safe set is represented by the black dashed lines, where the green shaded region is safe and the red shaded region is in violation of the constraint. Each ``x" represents the initial conditions of a test case. A green ``x" corresponds to no constraints being violated where as red represents the constraint being violated. From this analysis, 98.32\% of test cases were successful. Initial points are not generated up to the boundaries due to the $5\%$ buffer and the tuned biases. Additionally, all constraints, the HOCBFs and CBFs, are evaluated with the initial conditions. If any constraints are violated then a new initial condition is chosen. This is especially evident in the plots for the initial conditions of the spacecrafts angular velocity (\cref{sf:ic:w1,sf:ic:w2,sf:ic:w3}) and the initial condition for energy (\cref{sf:ic:en}) where the set of initial conditions either doesn't reach to the $5\%$ buffer or becomes sparser as it approaches. Since the slack variable is implemented on the communication constraint, the analysis does not consider this constraint to ever be violated. Of the simulations that failed $41.67\%$ failed because only one constraint was violated,  $34.52\%$ failed due to two constraints being violated,  $21.43\%$ failed because of three constraints being violated, and  $2.38\%$ failed because four constraints were violated. Of the constraints that were violated, the attitude exclusion zone was violated the most accounting for $51.19\%$ of the failed simulations. The second most prevalent violation was of the limits on $\psi_y$ which accounted for $47.62\%$ of the failed simulations. The constraints on $\psi_z$ and maintaining battery charge were violated in $40.48\%$ and $38.10\%$ of failed simulations, respectively. Lastly, $\psi_x$ and temperature constraints accounted for $4.76\%$ and $2.38\%$ of the failed simulations, respectively. Note, summing the percentage of failures caused by each constraint does not add up to $100\%$ since more than one constraint can be violated in a single simulation. While there are not specific values for certain initial conditions that result in simulation failures, it is observed from \cref{sf:ic:psi2,sf:ic:psi3} that initial values for $\psi_y$ and $\psi_z$ near their limits have a greater concentration of failures compared to the full range of values.  

\begin{figure*}[h!]
    \centering
    \subfigure[]{\includegraphics[width=0.3\textwidth]{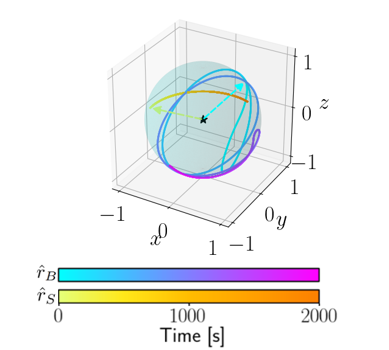}}
    \subfigure[]{\includegraphics[width=0.3\textwidth]{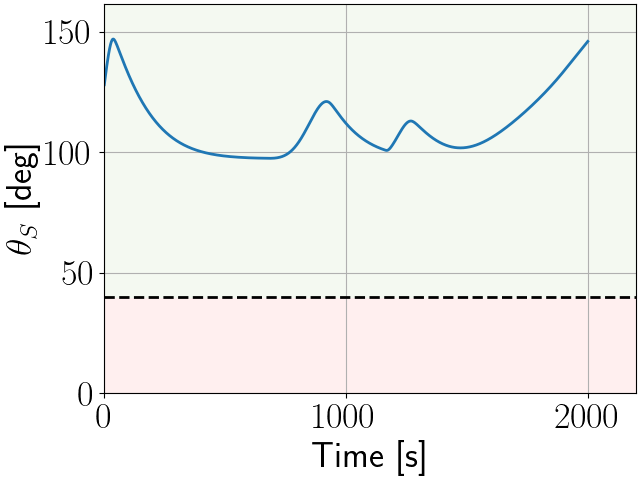}}
    \subfigure[]{\includegraphics[width=0.3\textwidth]{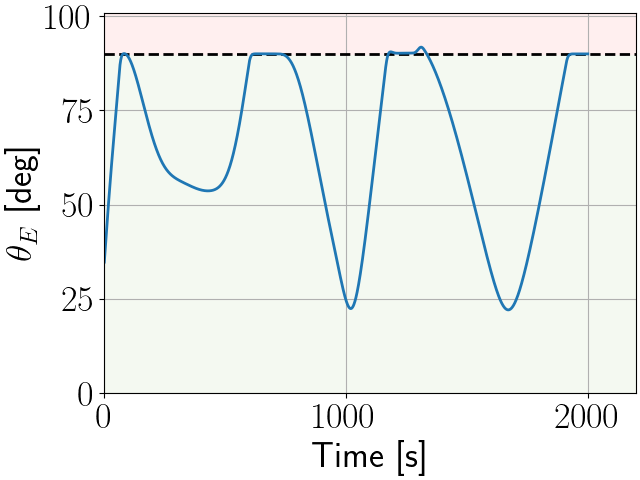}}
    
    \subfigure[]{\includegraphics[width=0.3\textwidth]{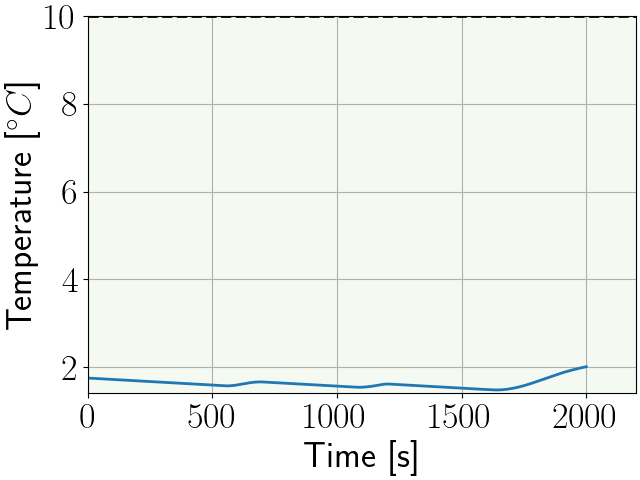}}
    \subfigure[]{\includegraphics[width=0.3\textwidth]{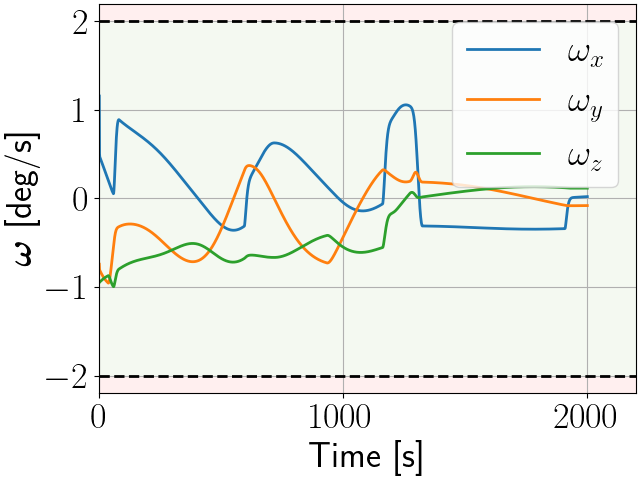}}
    \subfigure[]{\includegraphics[width=0.3\textwidth]{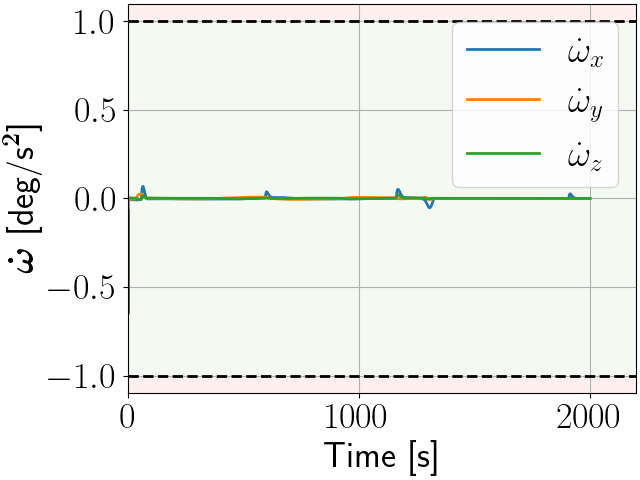}}
    
    \subfigure[]{\includegraphics[width=0.3\textwidth]{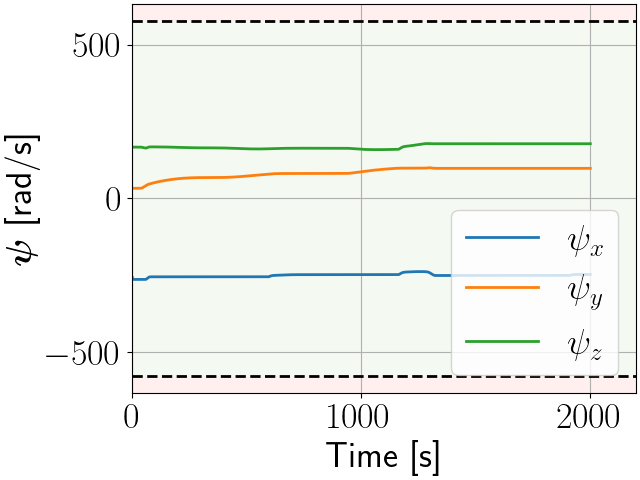}}
    \subfigure[]{\includegraphics[width=0.3\textwidth]{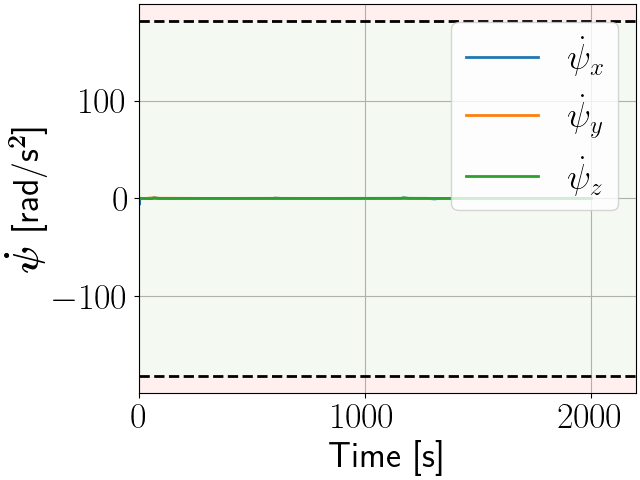}}
    \subfigure[]{\includegraphics[width=0.3\textwidth]{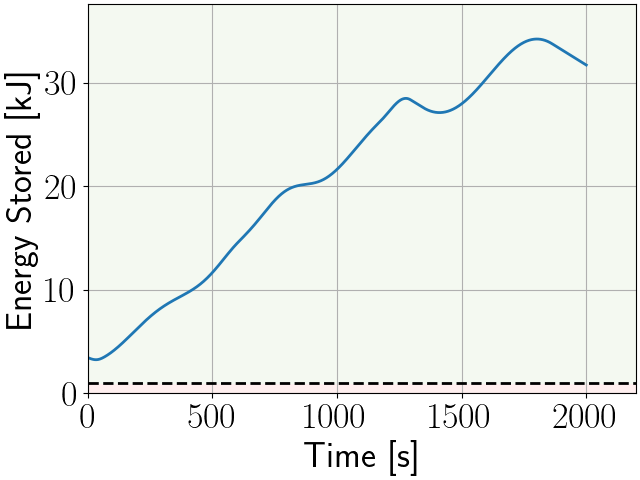}}
    \caption{Example of a simulation with RTA and slack variable on the communication constraint.}
    \label{f:sim-pass}
\end{figure*}

Figure~\ref{f:sim-pass} shows an example of a passed simulation. The initial conditions for this simulation are ${\bm{q}}=[0.404, -0.225, 0.406, 0.789], {\bm{\omega}}=[1.16,-0.742,-0.951] \,deg/s$, ${\bm{\psi}}=[-257,34,167] \, rad/s$, $T=1.75^o C$, $E=3.4 \,kJ$, and $\theta_{S} = 186^o$. While the communication constraint is violated at approximately 1100 seconds, likely due to prioritizing the angular velocity constraints, it is strictly enforced despite being close to the boundary at three other time intervals. This confirms that the slack variable enables a constraint to be violated only if necessary to prevent the QP from failing, while allowing all other constraints to still be enforced. 

\begin{figure*}[h!]
    \centering
    \subfigure[]{\includegraphics[width=0.3\textwidth]{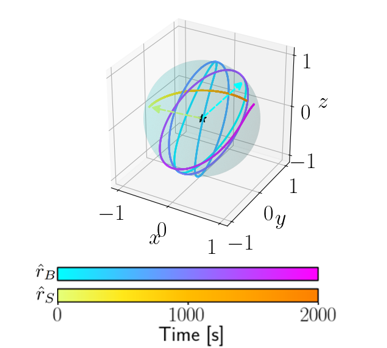}\label{f:f:vec}}
    \subfigure[]{\includegraphics[width=0.3\textwidth]{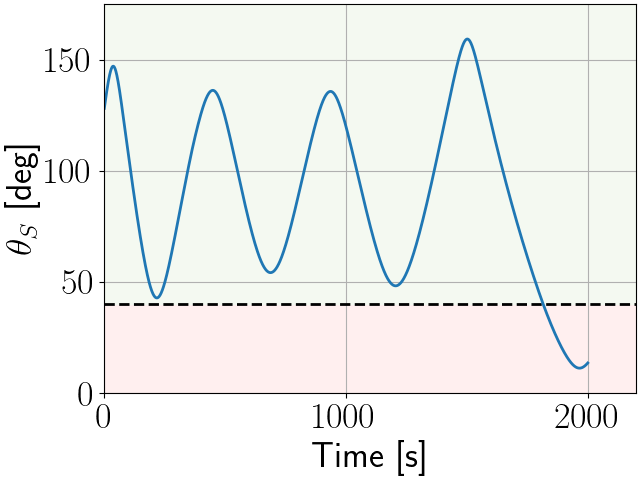}\label{f:f:koz_sun}}
    \subfigure[]{\includegraphics[width=0.3\textwidth]{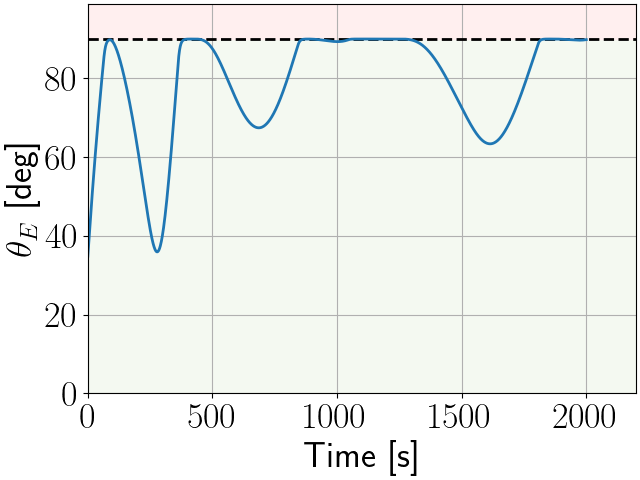}Figure~\label{f:f:koz_earth}}
    
    \subfigure[]{\includegraphics[width=0.3\textwidth]{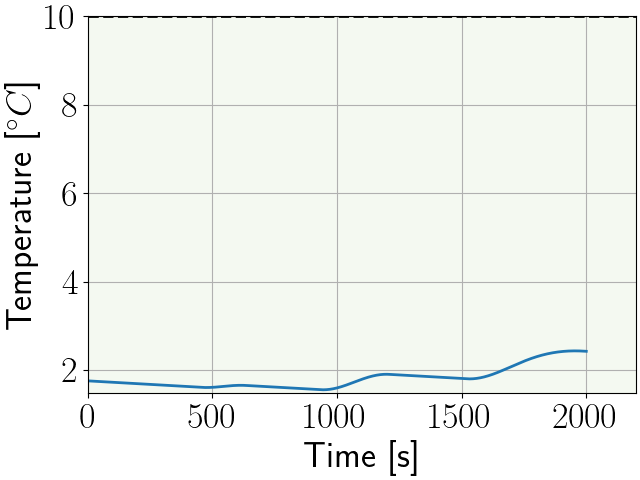}}
    \subfigure[]{\includegraphics[width=0.3\textwidth]{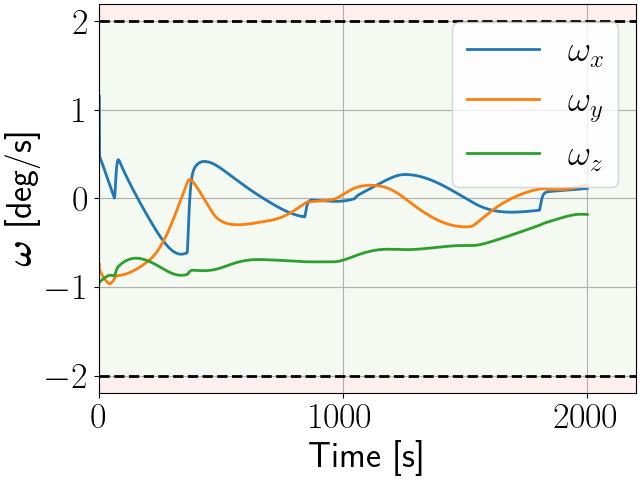}}
    \subfigure[]{\includegraphics[width=0.3\textwidth]{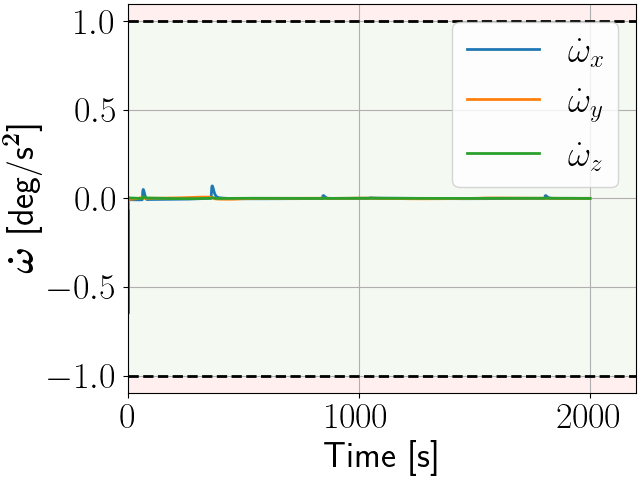}}
    
    \subfigure[]{\includegraphics[width=0.3\textwidth]{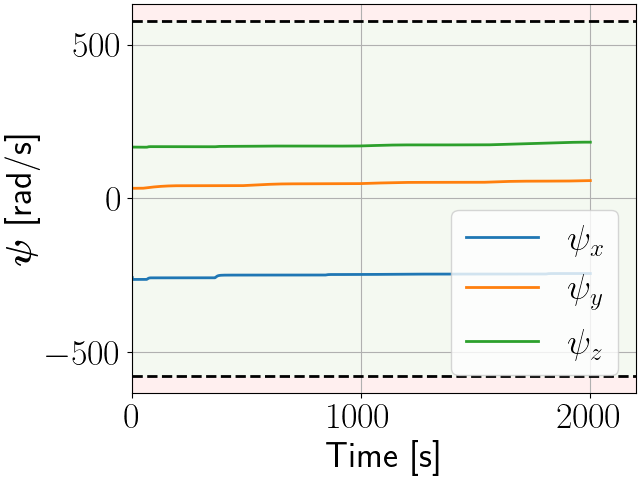}}
    \subfigure[]{\includegraphics[width=0.3\textwidth]{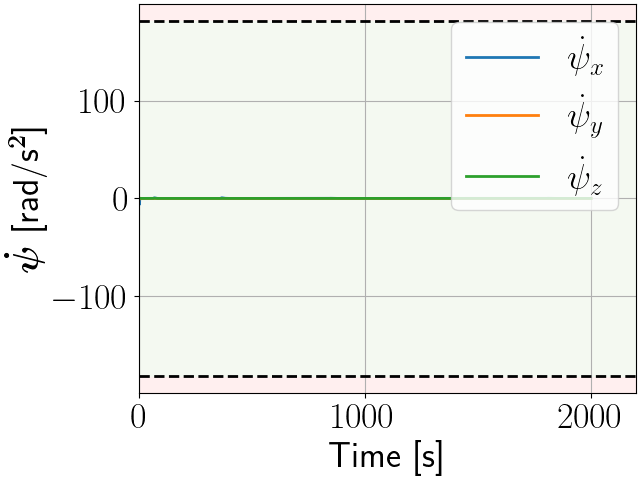}}
    \subfigure[]{\includegraphics[width=0.3\textwidth]{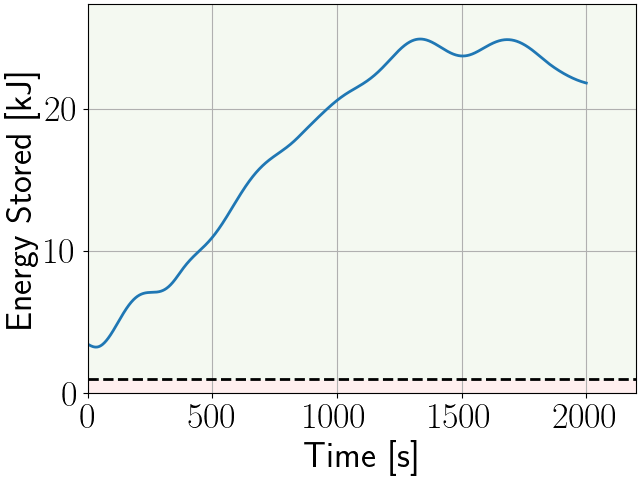}}
    \caption{Example of a simulation that failed and violated multiple constraints: attitude exclusion zone, communication, and maintain battery charge.}
    \label{f:sim-koz}
\end{figure*}

A slack variable can be implemented on a desired constraint to be violated when needed to prevent the QP from failing. Figure~\ref{f:sim-koz} shows an example of the slack variable incorporated into the attitude exclusion constraint rather than the communication constraint. The initial conditions are the same as those in Figure~\ref{f:sim-pass}. As seen in Figure~\ref{f:f:koz_sun}, at approximately 1800 seconds the attitude exclusion constraint is violated, but the communication constraint, shown in Figure~\ref{f:f:koz_earth} is not. This is expected since the slack is implemented on the attitude exclusion constraint. This demonstrates the prioritization capability of incorporating slack variables into constraints as desired.

\section{Conclusions} \label{s:Con}
This paper developed several safety constraints for spacecraft attitude maneuvering motivated by the autonomous spacecraft inspection problem. ASIF RTA, a minimally invasive optimization-based safety assurance technique, is able to enforce all constraints simultaneously with and without a primary controller active for the spacecraft. Slack variables were shown to be a method for constraint prioritization when all constraints cannot be satisfied. Monte-Carlo analysis showed that the ASIF RTA was able to enforce all constraints simultaneously for almost all test points. 

Future work in this topic includes: combining translational dynamics and constraints \cite{dunlap2023run2} for a full six degree-of-freedom spacecraft completing an inspection task with multiple deputies while ensuring safety through RTA, prioritization of constraints through slack variables on multiple constraints, and incorporating sensor uncertainty into RTA while still guaranteeing safety of the system.

\acknowledgments
This work is supported by the Department of Defense (DoD) through the Air Force Research Laboratory (AFRL) Seedings for Disruptive Capabilities Program (SDCP) called Safe Trusted Autonomy for Responsible Spacecraft (STARS). The views expressed are those of the authors and do not reflect the official guidance or position of the United States Government, the Department of Defense or of the United States Air Force.

\bibliographystyle{IEEEtran}
\bibliography{references}

\begin{thebibliography}{10}
\providecommand{\url}[1]{#1}
\csname url@samestyle\endcsname
\providecommand{\newblock}{\relax}
\providecommand{\bibinfo}[2]{#2}
\providecommand{\BIBentrySTDinterwordspacing}{\spaceskip=0pt\relax}
\providecommand{\BIBentryALTinterwordstretchfactor}{4}
\providecommand{\BIBentryALTinterwordspacing}{\spaceskip=\fontdimen2\font plus
\BIBentryALTinterwordstretchfactor\fontdimen3\font minus
  \fontdimen4\font\relax}
\providecommand{\BIBforeignlanguage}[2]{{%
\expandafter\ifx\csname l@#1\endcsname\relax
\typeout{** WARNING: IEEEtran.bst: No hyphenation pattern has been}%
\typeout{** loaded for the language `#1'. Using the pattern for}%
\typeout{** the default language instead.}%
\else
\language=\csname l@#1\endcsname
\fi
#2}}
\providecommand{\BIBdecl}{\relax}
\BIBdecl

\bibitem{OSOidx}
\BIBentryALTinterwordspacing
{United Nations Office for Outer Space Affairs}, ``Online index of objects
  launched into outer space,'' accessed: 2023-07-03. [Online]. Available:
  \url{https://www.unoosa.org/oosa/osoindex/search-ng.jspx}
\BIBentrySTDinterwordspacing

\bibitem{arney2021orbit}
D.~Arney, R.~Sutherland, J.~Mulvaney, D.~Steinkoenig, C.~Stockdale, and
  M.~Farley, ``On-orbit servicing, assembly, and manufacturing (osam) state of
  play,'' 2021.

\bibitem{hobbs2021risk}
K.~L. Hobbs, A.~R. Collins, and E.~M. Feron, ``Risk-based formal requirement
  elicitation for automatic spacecraft maneuvering,'' in \emph{AIAA Scitech
  2021 Forum}, 2021, p. 1122.

\bibitem{hobbs2023runtime}
K.~L. Hobbs, M.~L. Mote, M.~C. Abate, S.~D. Coogan, and E.~M. Feron, ``Runtime
  assurance for safety-critical systems: An introduction to safety filtering
  approaches for complex control systems,'' \emph{IEEE Control Systems
  Magazine}, vol.~43, no.~2, pp. 28--65, 2023.

\bibitem{hook2018initial}
L.~R. Hook, M.~Skoog, M.~Garland, W.~Ryan, D.~Sizoo, and J.~VanHoudt, ``Initial
  considerations of a multi-layered run time assurance approach to enable
  unpiloted aircraft,'' in \emph{2018 IEEE Aerospace Conference}.\hskip 1em
  plus 0.5em minus 0.4em\relax IEEE, 2018, pp. 1--11.

\bibitem{dunlap2023run}
K.~Dunlap, M.~Mote, K.~Delsing, and K.~L. Hobbs, ``Run time assured
  reinforcement learning for safe satellite docking,'' \emph{Journal of
  Aerospace Information Systems}, vol.~20, no.~1, pp. 25--36, 2023.

\bibitem{dunlap2023run2}
K.~Dunlap, D.~van Wijk, and K.~L. Hobbs, ``Run time assurance for autonomous
  spacecraft inspection,'' \emph{arXiv preprint arXiv:2302.02885}, 2023.

\bibitem{ames2016control}
A.~D. Ames, X.~Xu, J.~W. Grizzle, and P.~Tabuada, ``Control barrier function
  based quadratic programs for safety critical systems,'' \emph{IEEE
  Transactions on Automatic Control}, vol.~62, no.~8, pp. 3861--3876, 2016.

\bibitem{cortez2019control}
W.~S. Cortez, D.~Oetomo, C.~Manzie, and P.~Choong, ``Control barrier functions
  for mechanical systems: Theory and application to robotic grasping,''
  \emph{IEEE Transactions on Control Systems Technology}, vol.~29, no.~2, pp.
  530--545, 2019.

\bibitem{hsu2015control}
S.-C. Hsu, X.~Xu, and A.~D. Ames, ``Control barrier function based quadratic
  programs with application to bipedal robotic walking,'' in \emph{2015
  American Control Conference (ACC)}.\hskip 1em plus 0.5em minus 0.4em\relax
  IEEE, 2015, pp. 4542--4548.

\bibitem{dunlap2021comparing}
K.~Dunlap, M.~Hibbard, M.~Mote, and K.~Hobbs, ``Comparing run time assurance
  approaches for safe spacecraft docking,'' \emph{IEEE Control Systems
  Letters}, vol.~6, pp. 1849--1854, 2021.

\bibitem{wu2021attitude}
Y.-Y. Wu and H.-J. Sun, ``Attitude tracking control with constraints for rigid
  spacecraft based on control barrier lyapunov functions,'' \emph{IEEE
  Transactions on Aerospace and Electronic Systems}, vol.~58, no.~3, pp.
  2053--2062, 2021.

\bibitem{gurriet2018online}
T.~Gurriet, M.~Mote, A.~D. Ames, and E.~Feron, ``An online approach to active
  set invariance,'' in \emph{2018 IEEE Conference on Decision and Control
  (CDC)}.\hskip 1em plus 0.5em minus 0.4em\relax IEEE, 2018, pp. 3592--3599.

\bibitem{ames2019control}
A.~D. Ames, S.~Coogan, M.~Egerstedt, G.~Notomista, K.~Sreenath, and P.~Tabuada,
  ``Control barrier functions: Theory and applications,'' in \emph{2019 18th
  European control conference (ECC)}.\hskip 1em plus 0.5em minus 0.4em\relax
  IEEE, 2019, pp. 3420--3431.

\bibitem{nagumo1942lage}
M.~Nagumo, ``{\"U}ber die lage der integralkurven gew{\"o}hnlicher
  differentialgleichungen,'' \emph{Proceedings of the Physico-Mathematical
  Society of Japan. 3rd Series}, vol.~24, pp. 551--559, 1942.

\bibitem{xiao2022control}
W.~Xiao, C.~G. Cassandras, C.~A. Belta, and D.~Rus, ``Control barrier functions
  for systems with multiple control inputs,'' in \emph{2022 American Control
  Conference (ACC)}.\hskip 1em plus 0.5em minus 0.4em\relax IEEE, 2022, pp.
  2221--2226.

\bibitem{hill1878researches}
G.~W. Hill, ``Researches in the lunar theory,'' \emph{American journal of
  Mathematics}, vol.~1, no.~1, pp. 5--26, 1878.

\bibitem{markley2014fundamentals}
F.~L. Markley and J.~L. Crassidis, \emph{Fundamentals of spacecraft attitude
  determination and control}.\hskip 1em plus 0.5em minus 0.4em\relax Springer,
  2014, vol. 1286.

\bibitem{petersen2021challenge}
C.~D. Petersen, K.~Hobbs, K.~Lang, and S.~Phillips, ``Challenge problem:
  assured satellite proximity operations,'' in \emph{31st AAS/AIAA Space Flight
  Mechanics Meeting}, 2021.

\bibitem{foster2022small}
I.~Foster and {Air Force Research Laboratory}, ``Small satellite thermal
  modeling guide,'' 2022.

\bibitem{wertz1999space}
J.~R. Wertz, W.~J. Larson, D.~Kirkpatrick, and D.~Klungle, \emph{Space mission
  analysis and design}.\hskip 1em plus 0.5em minus 0.4em\relax Springer, 1999,
  vol.~8.

\bibitem{garzon2018thermal}
A.~Garz{\'o}n and Y.~A. Villanueva, ``Thermal analysis of satellite libertad 2:
  a guide to cubesat temperature prediction,'' \emph{Journal of Aerospace
  Technology and Management}, vol.~10, p. e4918, 2018.

\bibitem{bluePower}
\BIBentryALTinterwordspacing
``Power {S}ystems - {B}lue {C}anyon {T}echnologies,'' 2022, accessed:
  2023-07-13. [Online]. Available:
  \url{https://www.bluecanyontech.com/static/datasheet/BCT_DataSheet_Components_PowerSystems.pdf}
\BIBentrySTDinterwordspacing

\bibitem{ravaioli2023universal}
U.~J. Ravaioli, K.~Dunlap, and K.~Hobbs, ``A universal framework for
  generalized run time assurance with jax automatic differentiation,'' in
  \emph{2023 American Control Conference (ACC)}.\hskip 1em plus 0.5em minus
  0.4em\relax IEEE, 2023, pp. 4264--4269.

\end{thebibliography}

\thebiography

\begin{biographywithpic}{Cassie-Kay McQuinn}{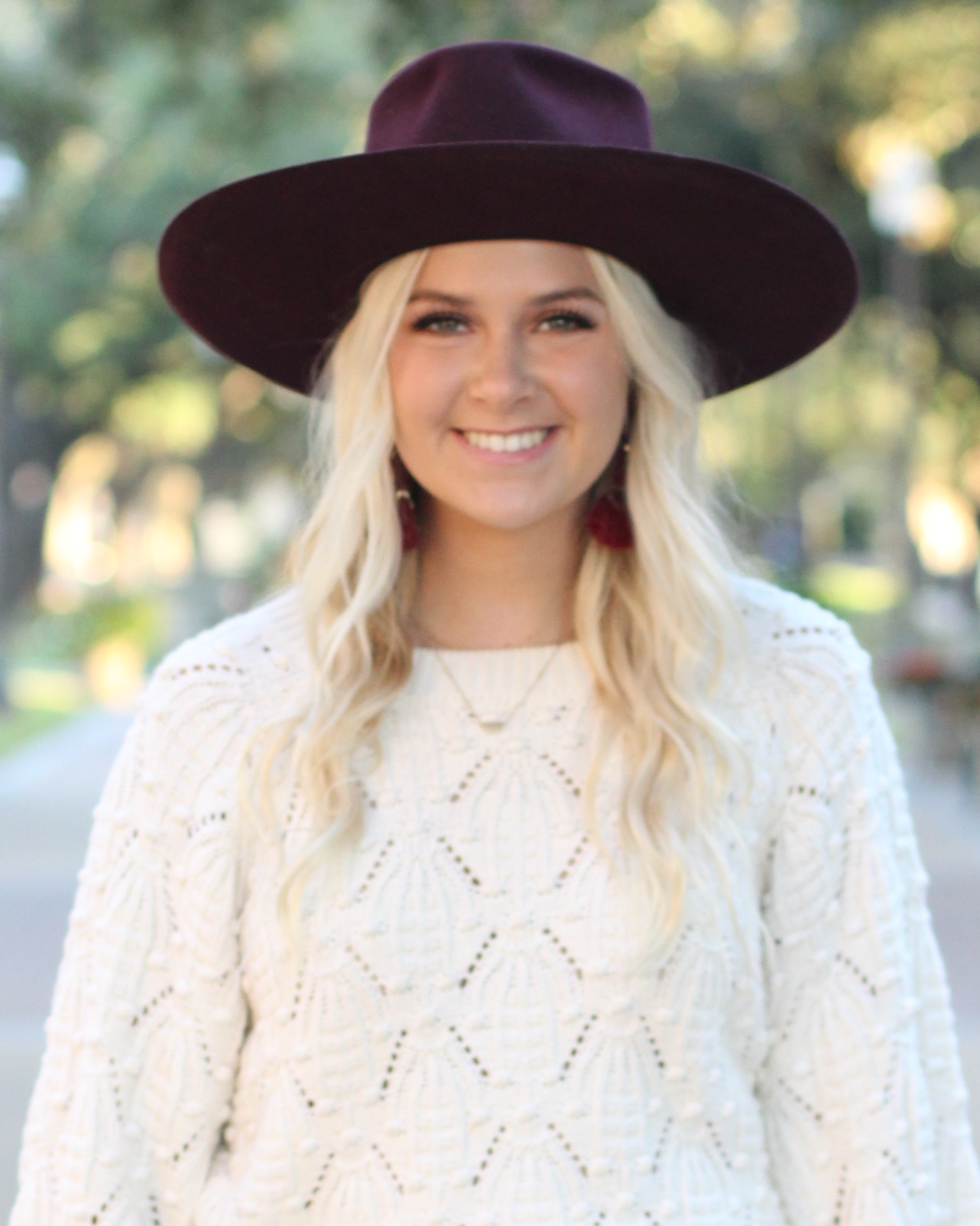}
    is a Master's student in Aerospace Engineering at Texas A\&M University. She received her B.S. in Aerospace Engineering from the same institution in 2021 with a minor in Mathematics and a certificate in Holistic Leadership. At Texas A\&M she is a graduate research assistant in the Vehicle Systems and Control Laboratory (VSCL), there her work focuses on aircraft system identification from flight test data. Her current research interests are in aerospace vehicle control and dynamics, incorporation of autonomy, and implementation of this work through flight testing on unmanned air vehicles (UAVs).
\end{biographywithpic}

\begin{biographywithpic}{Kyle Dunlap}{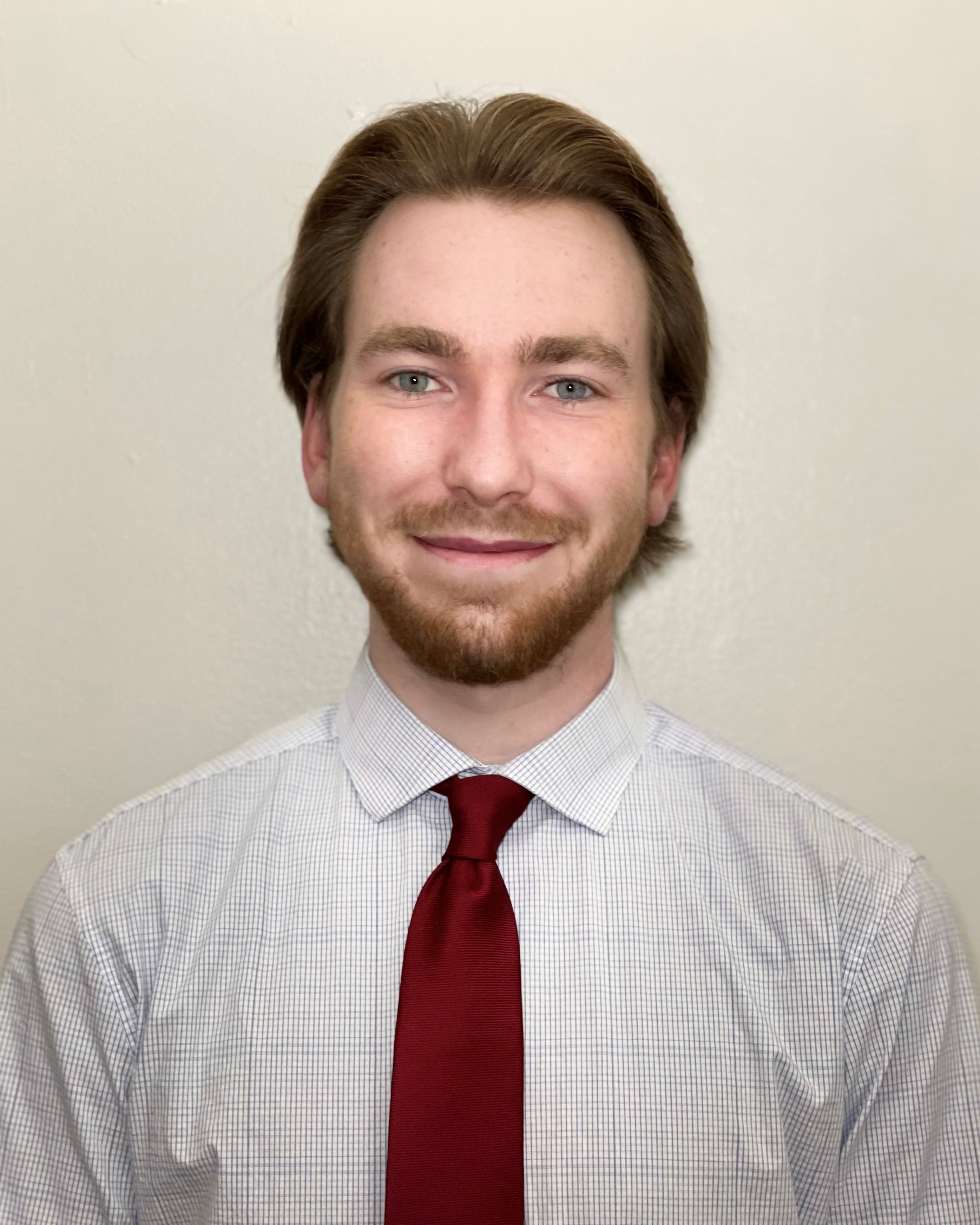}
is an AI Scientist at Parallax Advanced Research, primarily working with the Safe Autonomy Team at the Air Force Research Laboratory's Autonomy Capability Team (ACT3). There he investigates real time safety assurance methods for intelligent aerospace control systems. His previous experience includes developing a universal framework for Run Time Assurance (RTA) and comparing different RTA approaches during Reinforcement Learning training. He received his BS and MS in Aerospace Engineering from the University of Cincinnati.
\end{biographywithpic}

\begin{biographywithpic}{Nathaniel Hamilton}{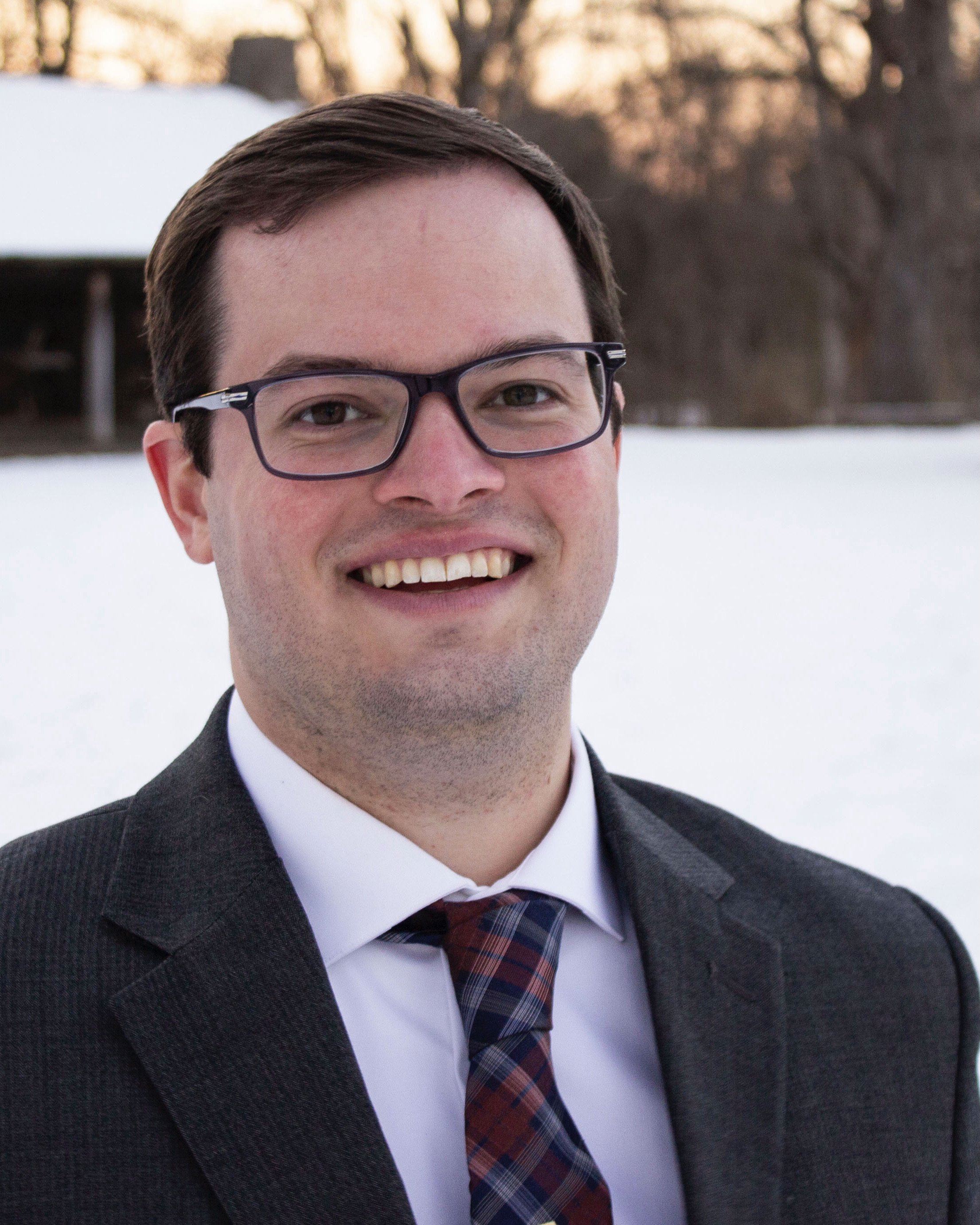}
    is an AI Scientist at Parallax Advanced Research who works closely with the Safe Autonomy Team at the Air Force Research Laboratory's Autonomy Capability Team (ACT3). There he investigates Safe Reinforcement Learning (SafeRL) approaches and how we can better integrate safety into the learning process to enable safe, trusted, and certifiable autonomous and learning-enabled controllers for aircraft and spacecraft applications. His previous experience includes studying how Run Time Assurance (RTA) impacts the learning and performance of SafeRL agents, and work in simulation to real-world (sim2real) transfer for learning-enabled controllers. In 2019, Dr. Hamilton was awarded the National Defense Science and Engineering Graduate (NDSEG) Fellowship. Dr. Hamilton has a BS in Electrical and Computer Engineering from Lipscomb University, and an MS and Ph.D. in Electrical Engineering from Vanderbilt University.
\end{biographywithpic}

\begin{biographywithpic}{Jabari Wilson}{Figures/Jabari}
is a PhD Student in Engineering Education at the University of Florida, hailing from Huntsville, AL. He received his bachelor's degree in mechanical engineering from The University of Alabama in 2019. Afterwards, he completed his master's degree in mechanical engineering with a minor in electrical engineering at University of Florida in 2022. Over the course of his academic career, he has completed 6 semesters of technical internship experience in a variety of areas, including project management, quality assurance, and quality control. These experiences were gained across the construction, automotive, and research industries. He has a passion for educational outreach and believes that the best work we can do is work that benefits the future of the world and those in it. So far, he has served as an assistant director for a Rockets and Robotics Summer Camp in his hometown and peer advisor in the Florida Board of Education Summer Fellowship Program. From 2022-2023 academic year, Mr. Wilson has served as both technical lead and mentor to high school junior and seniors in Washington D.C. public schools as part of their Career Technical Education (CTE) Advanced Internship Placement (AIP) and Career-Ready Internship (CRI) programs through his position as an Air Force Research Lab (AFRL) Scholar with the Air Force Office of Scientific Research (AFOSR). Mr. Wilson also spent the summer of 2023 working with the National Science Foundation’s Eddie Bernice Johnson INCLUDES Initiative team on portfolio analysis and program evaluation. 
\end{biographywithpic}

\begin{biographywithpic}{Kerianne L. Hobbs}{Figures/HobbsKerianne8x10}
is the Safe Autonomy and Space Lead on the Autonomy Capability Team (ACT3) at the Air Force Research Laboratory. There she investigates rigorous specification, analysis, bounding, and intervention techniques to enable safe, trusted, ethical, and certifiable autonomous and learning controllers for aircraft and spacecraft applications. Her previous experience includes work in automatic collision avoidance and autonomy verification and validation research. Dr. Hobbs was selected for the 2020 AFCEA 40 Under 40 award and was a member of the team that won the 2018 Collier Trophy (Automatic Ground Collision Avoidance System Team), as well as numerous AFRL Awards.  Dr. Hobbs has a BS in Aerospace Engineering from Embry-Riddle Aeronautical University, an MS in Astronautical Engineering from the Air Force Institute of Technology, and a Ph.D. in Aerospace Engineering from the Georgia Institute of Technology. 
\end{biographywithpic}

\end{document}